\documentclass[prb,twocolumn,superscriptaddress]{revtex4}

\usepackage{amsmath}
\usepackage{amsfonts}
\usepackage{amssymb,mathrsfs}
\usepackage{bbold,yfonts}
\usepackage{upgreek}
\usepackage{graphicx}
\usepackage{xcolor}

\newcommand{\bs}[1]{\boldsymbol{#1}}

\begin{document}

\title{Magnetohydrodynamics in graphene: Shear and Hall viscosities}

\author{B. N. Narozhny}
\affiliation{\mbox{Institut f\"ur Theorie der kondensierten Materie, Karlsruhe Institute of
Technology, 76128 Karlsruhe, Germany}}
\affiliation{National Research Nuclear University MEPhI (Moscow Engineering Physics Institute),
  115409 Moscow, Russia}
\author{M. Sch\"utt}
\affiliation{Condensed Matter Theory Group, Paul Scherrer Institute, 5232 Villigen PSI, Switzerland}
\affiliation{Department of Theoretical Physics, University of Geneva, 1211 Geneva, Switzerland}
\affiliation{Institute for Theoretical Physics, ETH Zurich, 8093 Zurich, Switzerland}

\date{\today}

\begin{abstract}
   Viscous phenomena are the hallmark of the hydrodynamic flow
   exhibited by Dirac fermions in clean graphene at high enough
   temperatures. We report a quantitative calculation of the
   electronic shear and Hall viscosities in graphene based on the
   kinetic theory combined with the renormalization group providing a
   unified description at arbitrary doping levels and nonquantizing
   magnetic fields. At charge neutrality, the Hall viscosity vanishes,
   while the field-dependent shear viscosity decays from its
   zero-field value saturating to a nonzero value in classically
   strong fields. Away from charge neutrality, the field-dependent
   viscosity coefficients tend to agree with the semiclassical
   expectation.
\end{abstract}

\maketitle

Hydrodynamic behavior of charge carriers in graphene has been the
focus of considerable experimental
\cite{ong,kim0,nlr,geim1,kim1,mac,kim2,geim2,geim3,ihn,goo,sulp,gal,geim4}
and theoretical efforts
\cite{har,kash,mfs,mfss,mus,job,alf,msf,pol15,hydro0,hydro1,fl0,poli16,lsk,fl1,fl2,ssk,lev17,lev18,adam,julia1,ks19,lev,bur19}
(recently reviewed in Refs.~\onlinecite{rev,luc}). Within linear
response, the difference between an Ohmic current and a hydrodynamic
flow is determined by the viscosity
\cite{geim1,geim2,geim4,fl0,fl2,pol15,hydro1} (see
Refs.~\onlinecite{geim3,sulp} on the issue of ballistic electrons).

In traditional hydrodynamics \cite{dau6}, viscosity is a measure of
mutual friction between the neighboring fluid elements moving with
distinct velocities. From the viewpoint of the microscopic theory,
viscosity is a fourth-rank tensor that can be defined as a ``response
function'' relating the stress (or momentum flux) to the strain
\cite{read,julia1}. In isotropic systems \cite{dau6}, the viscosity
tensor contains two independent coefficients, the shear and bulk
viscosities. The latter is only important for physical phenomena
associated with fluid compressibility and is known to vanish for
monoatomic gases \cite{dau6}, ultrarelativistic systems
\cite{kha,dau10}, and Fermi liquids \cite{akha}. Similarly, it's been
argued to vanish in graphene \cite{msf,hydro1,poli16}, at least to the
leading approximation. In anisotropic systems the situation is more
involved \cite{julia}.

Treating the shear viscosity as a linear response function, one can
derive a Kubo formula \cite{poli16,read,julia1} that can be related to
the Kubo formula for electrical conductivity \cite{read}. In practice,
Kubo-formula based calculations are typically perturbative and one can
only use this approach to evaluate the viscosity either in doped
graphene (i.e., in the so-called ``Fermi-liquid'' or ``degenerate''
limit) \cite{poli16} or in the high-frequency collisionless regime
\cite{julia1}. At charge neutrality, one typically turns to the
kinetic theory \cite{kash,msf} combined with the renormalization group
\cite{shsch}.

The relation between the Kubo formulas for shear viscosity and
conductivity \cite{read} leads to certain expectations regarding the
dependence of the viscosity tensor on the external magnetic field. In
the simplest case \cite{ale,moo,stein}, one finds the field-dependent
shear viscosity and the newly appearing Hall viscosity to mimic the
magnetoconductivity and Hall conductivity in the usual Drude theory,
respectively, with the only difference that the scattering time is now
provided by electron-electron interactions.

An experimental measurement of the electronic shear viscosity is a
nontrivial task \cite{corb}. Based on nonlocal resistance measurements
\cite{geim1}, a related quantity -- the kinematic viscosity
\cite{dau6} -- was estimated to have a ``higher-than-in-honey'' value
${\nu\approx0.1}\,$m$^2$/s at typical charge densities,
${n\sim10^{12}}\,$cm$^{-2}$, and temperatures, ${T=220}\,$K. This
value is of the same order of magnitude as the theoretical expectation
\cite{poli16} for doped graphene and agrees with more recent
measurements \cite{geim2,geim4}. In contrast, the theoretical result
for the shear viscosity in neutral graphene \cite{kash,msf,hydro1} has
not been tested experimentally. Hall viscosity has been studied in a
recent experiment reported in Ref.~\onlinecite{geim4}. Again, the
measurements were performed away from charge neutrality, where the
shear and Hall viscosities follow the semiclassical field dependence
\cite{ale,moo,pol17}.

The purpose of the present paper is to provide a consistent, unified
calculation of the shear and Hall viscosities in graphene at arbitrary
doping levels within the ``hydrodynamic'' temperature window
\cite{rev} and at arbitrary nonquantizing magnetic fields. In a
companion paper \cite{me1}, we have generalized the nonlinear
hydrodynamic equations derived in Ref.~\onlinecite{hydro1} to include
the external magnetic field. Here we report a quantitative calculation
of the kinetic coefficients on the basis of the kinetic theory
combined with the renormalization group approach. At neutrality
and in the degenerate regime, the results can be obtained in a
closed analytical form. For arbitrary carrier densities, the
viscosities can be expressed in terms of certain ``scattering rates''
to be evaluated numerically.

The results of our calculations are in good agreement with available
experimental evidence. In the absence of the magnetic field, we find
the kinematic viscosity to depend rather weakly on the carrier
density, remaining of the order of ${\nu\approx0.1}\,$m$^2$/s for all
densities explored in the experiment of Ref.~\onlinecite{geim1}. The
field-dependent shear and Hall viscosities away from charge neutrality
are qualitatively similar to the semiclassical expectations
\cite{geim4}, reaching the standard Drude-like dependence at the
experimental densities, ${n\sim10^{12}}\,$cm$^{-2}$. In contrast, the
shear viscosity at the neutrality point in classically strong magnetic
fields saturates to a nonzero value. Our results are illustrated in
physical units in Figs.~\ref{fig:nux} and \ref{fig:vhb}.

\section{From the microscopic theory to hydrodynamics}

We begin with a brief review of the hydrodynamic approach and a
discussion of the applicability of hydrodynamics to Dirac fermions in
graphene. The ideas presented here were developed in full detail in
Refs.~\onlinecite{dau10,har,hydro1,me1} (see the recent review
\cite{luc} for a more detailed discussion and a complete set of
references) and are included here to make the paper self-contained.

Hydrodynamics is a macroscopic manifestation of conservation laws in
an interacting many-body system that is a fluid. Typically
\cite{dau6}, one considers conservation of the particle number (or,
equivalently either mass or electric charge), energy, and
momentum. The latter provides the most stringent restrictions on the
applicability of hydrodynamics limiting one to systems with the only
(or the dominant) scattering mechanism being due to interparticle
collisions (e.g., electron-electron interaction) that conserve
momentum. At first glance this rules out scattering with other types
of excitations (e.g., electron-phonon or electron-magnon scattering)
as well as microscopically varying external potentials (e.g.,
potential disorder). In a typical solid, such processes dominate the
linear-response transport properties and while they can be accounted
for using the Boltzmann kinetic theory \cite{dau10}, a fully
hydrodynamic description is not possible.

In recent years, a number of materials have been developed
\cite{geim1,kim1,mac,ihn,goo} which are, on one hand, {\it clean
  enough}, such that disorder scattering is important only at the
lowest temperatures and on the other hand rigid enough, such that the
electron-phonon interaction is relevant at much higher
temperartures. This provides for a considerable intermediate
temperature range \cite{hydro1,rev} where the electron-electron
interaction is the dominant scattering mechanism in the
system. Mathematically, the above can be summarized by the inequality,
\begin{equation}
\label{con1}
\tau_{\rm ee} \ll \tau_{\rm dis}, \tau_{\rm e-ph}, \,\, {\rm etc},
\end{equation}
where $\tau_{\rm ee}$ is the typical time scale associated with the
electron-electron interaction, $\tau_{\rm dis}$ describes disorder
scattering, $\tau_{\rm e-ph}$ -- the electron-phonon interaction, and
``etc'' stands for any other scattering-related time scale in the
problem (e.g., the ``recombination'' time $\tau_R$, see below).

Assume that the condition (\ref{con1}) is sufficient to uphold the
conservation laws in the electronic system. Then they can be expressed
in terms of continuity equations, that can be either written down on
symmetry grounds \cite{dau6} or can be obtained by integrating the
kinetic equation \cite{dau10}. The first of these equations is the
standard continuity equation reflecting charge conservation
\begin{subequations}
\label{ces}
\begin{equation}
\label{cen1}
\partial_t n + \bs{\nabla}_{\bs{r}}\!\cdot\!\bs{j} = 0,
\end{equation}
which in this paper we express in terms of the carrier density $n$ and
current $\bs{j}$, such that the actual charge density and electric
current are obtained by multiplying these quantities by the electron
charge.

The second equation reflects energy conservation. In the case of
charge carriers, this equation differs \cite{rev,luc} from its
textbook counterpart \cite{dau6,dau10} by an extra term describing
Joule's heat
\begin{equation}
\label{cene1}
\partial_t n_E + \bs{\nabla}_{\bs{r}}\!\cdot\!\bs{j}_E = e \bs{E}\cdot\bs{j}.
\end{equation}
Here $n_E$ and $\bs{j}_E$ are the energy density and current,
$\bs{E}$ is the electric field, and $e$ is the electron charge.

The third equation describes momentum conservation. In contrast to the
corresponding equation for a neutral fluid \cite{dau6,dau10}, this
equation takes into account the effect of electromagnetic
fields. Moreover, for reasons that will become clear below, we include
a small [as required by Eq.~(\ref{con1})] disorder-scattering term
\cite{hydro0,hydro1}
\begin{equation}
\label{cek1}
\partial_t n^\alpha_{\bs{k}} + \nabla^\beta_{\bs{r}} \Pi_E^{\alpha\beta}
- e n E^\alpha - \frac{e}{c} \left[\bs{j}\!\times\!\bs{B}\right]^\alpha =
- \frac{n^\alpha_{\bs{k}}}{\tau_{\rm dis}}.
\end{equation}
Here $\bs{n}_{\bs{k}}$ is the momentum density, $\Pi_E^{\alpha\beta}$
is the momentum flux (or stress tensor), and $\bs{B}$ is the magnetic
field.

The continuity equations (\ref{cen1})~-~(\ref{cek1}) are valid for any
electronic system satisfying Eq.~(\ref{con1}). In the particular case
of graphene, one may neglect scattering processes that change the
number of particles in each individual band (e.g., the Auger
processes, three-particle collisions, electron-phonon interaction,
etc.) such that the number of particles in each band should be
conserved separately. As a result, one finds another continuity
equation \cite{alf,rev,luc}
\begin{equation}
\label{ceni1}
\partial_t n_I + \bs{\nabla}_{\bs{r}}\!\cdot\!\bs{j}_I = -
\frac{n_I\!-\!n_{I,0}}{\tau_R},
\end{equation}
\end{subequations}
where $n_I$ and $\bs{j}_I$ are the so-called ``imbalance'' (or total
quasiparticle) density and currents that are related to the particle
number densities, $n_\pm$, and currents, $\bs{j}_\pm$, in the two
bands as
\begin{subequations}
\label{defs}
\begin{equation}
\label{ds}
n=n_+-n_-, \qquad n_I = n_++n_-,
\end{equation}
\begin{equation}
\bs{j} = \bs{j}_+ - \bs{j}_-,
\qquad
\bs{j}_I = \bs{j}_+ + \bs{j}_-.
\end{equation}
The right-hand side in Eq.~(\ref{ceni1}) describes weak [again, in the
  sense of Eq.~(\ref{con1})] quasiparticle recombination characterized
by a (long) time scale $\tau_R$. The quantity $n_{I,0}$ is the
equilibrium density of quasiparticles.

The kinetic derivation \cite{dau10,hydro1} of the continuity equations
(\ref{ces}) has the advantage of providing ``microscopic''
definitions of all macroscopic quantities in Eqs.~(\ref{ces}) in terms
of the quasiparticle distribution function. In graphene (or any other
two-band system), one may label the single-particle (band) states by
the band index, ${\lambda=\pm}$, and the momentum, $\bs{k}$. Denoting
the distribution function by $f_{\lambda\bs{k}}$, we define the above
densities and currents as 
\begin{equation}
\label{dpm}
n_+\!=\!N\!\!\int\!\!\frac{d^2k}{(2\pi)^2} f_{+,\bs{k}},
\quad
n_-\!=\!N\!\!\int\!\!\frac{d^2k}{(2\pi)^2} \left(1\!-\!f_{-,\bs{k}}\right),
\end{equation} 
\begin{equation}
\bs{j}\!=\!N\!\sum_\lambda\!\int\!\!\frac{d^2k}{(2\pi)^2} \bs{v}_{\lambda\bs{k}}f_{\lambda\bs{k}} ,
\quad
\bs{j}_I\!=\!N\!\sum_\lambda\!\int\!\!\frac{d^2k}{(2\pi)^2} \lambda \bs{v}_{\lambda\bs{k}}f_{\lambda\bs{k}},
\end{equation}
\begin{equation}
n_E = N\!\int\!\!\frac{d^2k}{(2\pi)^2} 
\left[ \epsilon_{+,\bs{k}} f_{+,\bs{k}} + \epsilon_{-,\bs{k}} \left(f_{-,\bs{k}}\!-\!1\right)\right],
\end{equation}
\begin{equation}
\label{jedef}
\bs{j}_E = N\!\sum_\lambda\!\int\!\frac{d^2k}{(2\pi)^2}
\epsilon_{\lambda\bs{k}} \bs{v}_{\lambda\bs{k}} f_{\lambda\bs{k}} = v_g^2\bs{n}_{\bs{k}},
\end{equation}
\begin{equation}
\label{nkdef}
\bs{n}_{\bs{k}} = N\!\sum_\lambda\!\int\!\frac{d^2k}{(2\pi)^2} \bs{k} f_{\lambda\bs{k}},
\end{equation}
\begin{equation}
\label{pe}
\Pi_E^{\alpha\beta} = N\!\sum_\lambda\!\int\!\!\frac{d^2k}{(2\pi)^2} 
k^\alpha v^\beta_{\lambda\bs{k}} f_{\lambda\bs{k}}.
\end{equation}
\end{subequations}
Here ${N=4}$ is the degeneracy factor. The second equality in
Eq.~(\ref{jedef}) is specific to the Dirac spectrum in graphene and
represents a crucial difference between the electron fluid in graphene
and the usual (massive) fluids. Indeed, assuming the Dirac form of the
quasiparticle spectrum \cite{kats}
\begin{equation}
\label{eg}
\epsilon_{\lambda\bs{k}} = \lambda v_g k,
\end{equation}
one immediately finds the following relations between velocity and
momentum (where $\bs{e}$ denotes the unit vector in the direction
indicated by the subscript)
\begin{equation}
\label{vg}
\bs{v}_{\lambda\bs{k}}=\lambda v_g \frac{\bs{k}}{k}, \quad
\bs{e}_{\bs{v}} = \bs{e}_{\bs{k}},
\quad
\bs{k} = \frac{\lambda k}{v_g}\bs{v}_{\lambda\bs{k}}
=\frac{\epsilon_{\lambda\bs{k}}\bs{v}_{\lambda\bs{k}}}{v_g^2}.
\end{equation}
Inserting Eqs.~(\ref{vg}) into Eqs.~(\ref{nkdef}) and (\ref{jedef}),
one concludes that (i) the momentum density in graphene is equivalent
to the energy flux, and (ii) the hydrodynamic flow in graphene
describes the energy flow in contrast to the standard hydrodynamics
describing the mass flow of a conventional fluid. Moreover,
conservation of momentum leads to the conclusion that the energy flux
in graphene is not relaxed by electron-electron interaction. The
system can only reach the steady state by means of (weak) disorder
scattering \cite{hydro0,hydro1} described by the last term in
Eq.~(\ref{cek1}).

\section{Ideal hydrodynamics in graphene}

The true equilibrium state (in the absence of the external fields) is
described by the Fermi-Dirac distribution function yielding constant
densities and zero currents, such that each term in Eqs.~(\ref{ces})
vanishes. Applying weak external fields one drives the system weakly
out of equilibrium. This can be described either by means of the
perturbative linear-response theory \cite{hydro0}, or, if the
condition (\ref{con1}) is fulfilled, by the hydrodynamic theory
\cite{rev,luc}. The latter approach requires two additional
assumptions.

First, one assumes the spatial and temporal variations of the external
fields and the resulting currents and density inhomogeneities to be
small [ultimately, in the sense of Eq.~(\ref{con1}) or the equivalent
  relation of the corresponding length scales], such that the
electron-electron scattering processes may maintain {\it local
  equilibrium}. The latter is described by the distribution function
\cite{har,hydro1,rev,luc,me1}
\begin{equation}
\label{le}
f^{(0)}_{\lambda\bs{k}} (\bs{r}) =
\left\{
1+\exp\left[\frac{\epsilon_{\lambda\bs{k}}\!-\!\mu_\lambda(\bs{r}) \!-\! 
\bs{u}(\bs{r})\!\cdot\!\bs{k}}{T(\bs{r})}\right]
\right\}^{-1},
\end{equation}
where ${\mu_\lambda(\bs{r})}$, $T(\bs{r})$, and $\bs{u}(\bs{r})$ are
the local chemical potential, local temperature, and hydrodynamic (or
``drift'') velocity, respectively.

Using the distribution function (\ref{defs}), one finds the
expressions for the equilibrium hydrodynamic quantities as well as the
thermodynamic pressure, $P$, and enthalpy, $W$, listed in
Appendix~\ref{leqs}. Substituting these quantities into the continuity
equations (\ref{ces}), one obtains the (ideal) hydrodynamic
equations. In particular, expressing the continuity equation
(\ref{cek1}) for the momentum density in terms of the hydrodynamic
velocity, $\bs{u}$, we obtain the generalized Euler equation
\cite{dau6}
\begin{eqnarray}
\label{eq0g}
&&
\!\!\!\!\!\!
W(\partial_t+\bs{u}\!\cdot\!\bs{\nabla})\bs{u}
+
v_g^2 \bs{\nabla} P
+
\bs{u} \partial_t P 
+
e(\bs{E}\!\cdot\!\bs{j})\bs{u} 
=
\\
&&
\nonumber\\
&&
\qquad\qquad\qquad\qquad\qquad
=
v_g^2 en\bs{E}
+
v_g^2 \frac{e}{c} \bs{j}\!\times\!\bs{B}
-
\frac{W\bs{u}}{\tau_{{\rm dis}}}.
\nonumber
\end{eqnarray}
Equation (\ref{eq0g}) was suggested in Ref.~\onlinecite{msf} in the
absence of electromagnetic fields and weak disorder (terms due to the
electric field were discussed in Ref.~\onlinecite{hydro1}). In
comparison to the standard Euler equation, the generalized equation
(\ref{eq0g}) contains two extra terms: (i) the time derivative of
pressure that can be interpreted as a reminder of the relativistic
nature \cite{har,luc} of the quasiparticle spectrum in graphene,
Eq.~(\ref{eg}), and (ii) the disorder scattering term necessary to
establish a steady state \cite{hydro0,hydro1,me1}.

Away from charge neutrality, the Euler equation (\ref{eq0g}) allows
for a homogeneous, steady flow \cite{hydro0} that is equivalent to the
usual Ohmic current [using Eqs.~(\ref{n0s}) - (\ref{hqsg}) with
  ${\mu_\pm=\mu}$ and ${\mu\gg{T}}$]
\[
v_g^2 en\bs{E}
+
v_g^2 \frac{e}{c} \bs{j}\!\times\!\bs{B}
=
\frac{\mu\bs{j}}{\tau_{{\rm dis}}},
\]
characterized by the standard Drude-like expressions for the
longitudinal and Hall resistivities 
\[
\rho_{xx}^{(0)} = \frac{\pi}{e^2|\mu|\tau_{\rm dis}},
\qquad
\rho_{xy}^{(0)} = R_H B,
\quad
R_H = - \frac{{\rm sign}\,\mu}{n|e|c},
\]
where $R_H$ is the Hall coefficient.

The complete set of the equations of ideal hydrodynamics in graphene
includes the generalized Euler equation (\ref{eq0g}), the continuity
equations (\ref{cen1}), (\ref{cene1}), and (\ref{ceni1}), as well as
the Poisson's equation relating the charge density to the electric
field and the (thermodynamic) equation of state \cite{har,hydro1}, see
Eqs.~(\ref{hqsg}),
\begin{equation}
\label{eqsta}
W=n_{E}+P = \frac{3 n_{E}}{2+u^2/v_g^2}.
\end{equation}
This equation has the thermodynamic nature and represents the second
assumption needed to build the hydrodynamic theory, namely, that
thermodynamic quantities and their relations are not affected by
the dissipative corrections to the ideal hydrodynamics \cite{dau10}.

\section{Generalized Navier-Stokes equation in graphene}

Taking into account dissipative processes modifies the macroscopic
quantities and turns the ideal Euler equation into the Navier-Stokes
equation, the central equation of the hydrodynamic theory \cite{dau6}.
Following the standard approach \cite{dau6,dau10,rev,luc}, we focus on
the velocity-independent kinetic coefficients. In graphene, these
include the viscosity and ``quantum conductivity'' (cf. the thermal
conductivity in the traditional hydrodynamics \cite{dau6}).

In the usual hydrodynamics \cite{dau6,hydro1,luc,me1}, viscosity is
defined as the coefficient in the leading term of the gradient
expansion of the dissipative correction to the stress tensor
\begin{subequations}
\label{dpdef}
\begin{equation}
\Pi_{E}^{\alpha\beta}=\Pi_{E,0}^{\alpha\beta}+\delta\Pi_{E}^{\alpha\beta},
\end{equation}
\begin{equation}
\label{visdefB}
\delta\Pi^{\alpha\beta}_E=-\eta(B)\textswab{D}^{\alpha\beta}
+\eta_H(B) \epsilon^{\alpha ij} \textswab{D}^{i\beta} e_B^j,
\end{equation}
\begin{equation}
\textswab{D}^{\alpha\beta}=
\nabla^\alpha u^\beta+\nabla^\beta u^\alpha-\delta^{\alpha\beta}\bs{\nabla}\!\cdot\!\bs{u},
\end{equation}
\end{subequations}
where ${\bs{e}_B=\bs{B}/B}$, $\eta(B)$, and $\eta_H(B)$ are the field
dependent shear \cite{ale,moo,stein} and Hall
\cite{ale,moo,stein,read,pol17,me1} viscosities (the latter appears
only in the presence of magnetic field; bulk viscosity in graphene
vanishes, at least to the leading approximation
\cite{hydro1,rev,luc}). While the sign of $\eta$ is fixed by
thermodynamics \cite{dau6,dau10}, the sign of $\eta_H$ is
not. Similarly to Ref.~\onlinecite{geim4}, we adopt the convention
where the Hall viscosity is positive for electrons \cite{me1} (and
negative for holes).

Substituting Eqs.~(\ref{dpdef}) into the continuity equation
(\ref{cek1}) and repeating the steps leading to the Euler equation
(\ref{eq0g}), we obtain the generalized Navier-Stokes equation
\cite{me1}
\begin{eqnarray}
\label{eq1g}
&&
\!\!\!\!\!\!
W(\partial_t+\bs{u}\!\cdot\!\bs{\nabla})\bs{u}
+
v_g^2 \bs{\nabla} P
+
\bs{u} \partial_t P 
+
e(\bs{E}\!\cdot\!\bs{j})\bs{u} 
=
\\
&&
\nonumber\\
&&
=
v_g^2 
\left[
\eta \Delta\bs{u}
-
\eta_H \Delta\bs{u}\!\times\!\bs{e}_B
+
en\bs{E}
+
\frac{e}{c} \bs{j}\!\times\!\bs{B}
\right]
-
\frac{\bs{j}_E}{\tau_{{\rm dis}}}.
\nonumber
\end{eqnarray}

Comparing the first terms in the left- and
right-hand sides of Eq.~(\ref{eq1g}), we define the kinematic
viscosity
\begin{equation}
\label{kv}
\nu = v_g^2\eta/W,
\end{equation}
with the dimensionality of the diffusion constant (m$^2$/s).

The second set of the kinetic coefficients describes the dissipative
corrections to the quasiparticle currents \cite{hydro1,me1} with
${\bs{F}_{\bs{u}}=e\bs{E}+(e/c)\bs{u}\!\times\!\bs{B}}$ (in the presence
of disorder, the energy current acquires a dissipative correction of
its own):
\begin{subequations}
\label{djdef}
\begin{equation}
\bs{j} = n \bs{u} + \delta\bs{j},
\quad
\bs{j}_I = n_{I} \bs{u} + \delta\bs{j}_I,
\quad
\bs{j}_{E} = W\bs{u} + \delta\bs{j}_E,
\end{equation}
\begin{equation}
\label{djs}
\begin{pmatrix}
  \delta\bs{j} \cr
  \delta\bs{j}_I \cr
  \delta\bs{j}_E/T
\end{pmatrix}
=
\widehat\Sigma\!
\begin{pmatrix}
\bs{F}_{\bs{u}}\!-\!T\,\bs{\nabla}\frac{\mu}{T} \cr
T\,\bs{\nabla}\frac{\mu_I}{T} \cr
0
\end{pmatrix}
+
\widehat\Sigma_H\!
\begin{pmatrix}
\bs{F}_{\bs{u}}\!-\!T\,\bs{\nabla}\frac{\mu}{T} \cr
T\,\bs{\nabla}\frac{\mu_I}{T} \cr
0
\end{pmatrix}\!\times\bs{e}_B,
\end{equation}
\begin{equation}
\label{mu}
\mu=(\mu_++\mu_-)/2, \qquad
\mu_I=(\mu_+-\mu_-)/2.
\end{equation}
\end{subequations}
The ``imbalance'' chemical potential $\mu_I$ is relevant for
thermoelectric effects \cite{alf}, which will be considered
elsewhere. In this paper we disregard the possibility of the
temperature gradients and set $\mu_\pm=\mu$ (or $\mu_I=0$).

At charge neutrality, the matrix $\widehat\Sigma$ is block diagonal,
i.e., the electric current decouples from the energy and imbalance
currents. For ${n=0}$, the total current $\bs{j}$ is given by the
dissipative correction $\delta\bs{j}$ which remains finite even in the
absence of disorder, $\tau_{\rm dis}\rightarrow\infty$,
\[
e\delta\bs{j} (\mu\!=\!0) = \sigma_Q\bs{E},
\qquad
\sigma_Q = {\cal A}e^2/\alpha_g^2,
\qquad
{\cal A}\approx0.12.
\]
Here $\sigma_Q$ is known as the ``quantum'' or ``intrinsic''
conductivity of graphene \cite{kash,mfs,schutt,hydro1,luc,me1}. Within
the electronic hydrodynamics in graphene, this quantity appears
instead of the usual thermal conductivity due to the special relation
between the energy current and momentum, see Eq.~(\ref{jedef}) and the
discussion following Eq.~(\ref{vg}).

Together with the continuity equations (\ref{ces}) and the equation of
state (\ref{eqsta}), the generalized Navier-Stokes equation
(\ref{eq1g}) forms a closed system of hydrodynamic equations that can
be solved in arbitrary geometries (see Ref.~\onlinecite{ks19} for the
appropriate boundary conditions). Away from charge neutrality, these
equations have to be solved together with the electrostatics equations
similarly to the usual Vlasov self-consistency \cite{dau10,hydro1}.
In free (e.g., suspended) graphene the latter is given by the
Poisson's equation. In gated structures used in the majority of
experiments \cite{geim1,geim2,geim3,geim4} the electrostatics is
dictated by the gate \cite{ash,mr1}, simplifying the relation between
the charge density and the electric field.

\section{Kinetic calculation of electronic viscosity in graphene}

In this section we report the kinetic theory results for the
electronic viscosity in graphene. The calculation method was outlined
in Ref.~\onlinecite{hydro1}, but unfortunately involves some tedious
algebra yielding the viscosity coefficients in terms of rather
cumbersome multidimensional integrals, see Appendix~\ref{visc}.  In
the simple limiting cases (e.g., at charge neutrality and in the
degenerate, ``Fermi-liquid'' regime), analytical results can be
obtained. Otherwise, for arbitrary doping levels the shear and Hall
viscosities, see Eqs.~(\ref{etares}), are computed numerically. These
results are illustrated in Figs.~\ref{fig:vb0} -
\ref{fig:etaHfl}. Details of the derivation are published in
Ref.~\onlinecite{me1}.

The kinetic theory is formally valid only in the weak coupling
limit. In particular, the ``collinear'' (or ``three-mode'')
approximation \cite{hydro1,mfss,me1} that allows us to solve the
kinetic equation is formally justified in the limit
${|\ln\alpha_g|\gg1}$ ($\alpha_g$ is the coupling constant in
graphene). Therefore, in order to obtain experimentally relevant
numerical values for the viscosity coefficients we supplement the
kinetic theory calculation with the renormalization group analysis
described in the following section.

\subsection{Shear viscosity at charge neutrality}

At charge neutrality, the general expressions (\ref{etares}) simplify
and can be evaluated analytically \cite{msf,hydro1} (up to a
multiplicative numerical factor). The Hall viscosity vanishes
identically (due to the exact electron-hole symmetry), while the shear
viscosity exhibits the following behavior.

In zero magnetic field, we recover the well-known result
\cite{msf} (the parametric dependence follows from the fact that the
only energy scale in neutral graphene is $T$)
\begin{equation}
\label{vdp}
\eta(B\!=\!0; \mu\!=\!0) = {\cal B} \frac{T^2}{\alpha_g^2 v_g^2}.
\end{equation}
The numerical coefficient was first evaluated in Ref.~\onlinecite{msf}
reporting the value ${{\cal B}=0.45}$. Although not explicitly
discussed \cite{msf}, this result was obtained using the ``bare''
Coulomb interaction, i.e., neglecting screening effects. Such an
approach is formally valid for asymptotically low temperatures
\cite{mfss,kash,schutt} where the coupling constant $\alpha_g$ is
expected to have been renormalized to a small enough value (see below
for the discussion of the renormalization group approach). Indeed,
evaluating the integrals (\ref{ttauij}) for the ``scattering rates''
with unscreened Coulomb interaction numerically, we find
${{\cal{B}}=0.446\pm0.005}$, where the deviation stems from the
systematic differences between various numerical methods. The small
difference between the above result and that of Ref.~\onlinecite{msf}
is due to the fact that our calculation takes into account only the
direct interaction, while the exchange term is small
\cite{kash,schutt} in $1/N$.

\begin{figure}[t]
\centerline{\includegraphics[width=0.9\columnwidth]{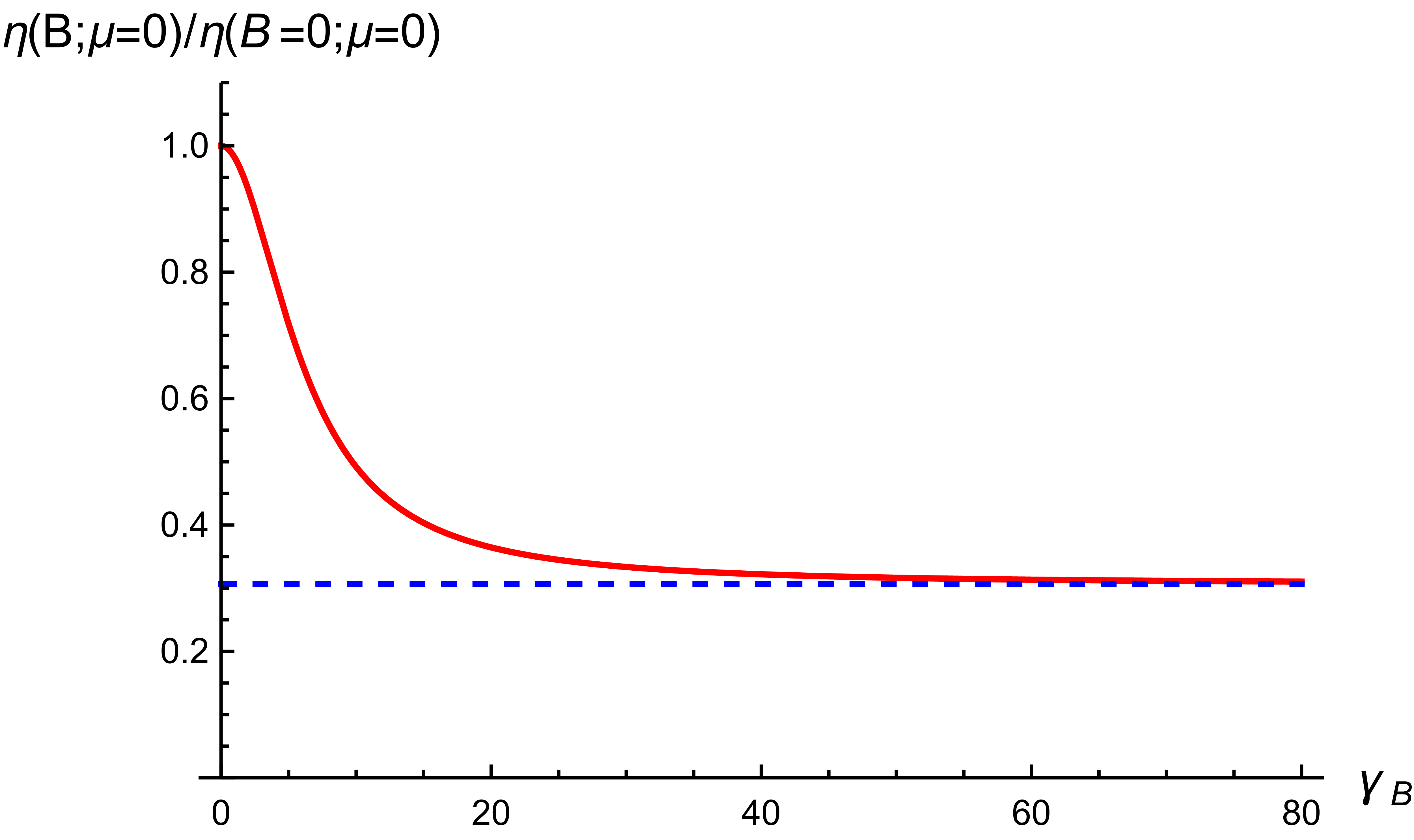}
}
\caption{Shear viscosity in graphene at charge neutrality. The solid
  red curve shows the ratio of the field-dependent viscosity to its
  zero-field value, ${\eta(B;\mu=0)/\eta(B=0;\mu=0)}$, as a function
  of $\gamma_B$, see Eqs.~(\ref{vdpb}) and (\ref{gammab}). In
  classically strong fields the viscosity saturates to
  ${\eta/\eta(B=0)={\cal B}_1/({\cal B}{\cal B}_2)\approx 0.3065}$
  (for unscreened interaction) shown by the blue dashed line.}
\label{fig:vb0}
\end{figure}

In magnetic field, the shear viscosity decreases but remains finite
in classically strong fields (due to the aforementioned decoupling of
the charge mode)
\begin{equation}
\label{vdpb}
\eta(B; \mu=0) = 
\frac{T^2}{\alpha_g^2v_g^2}
\frac{{\cal B}+{\cal B}_1\gamma_B^2}{1+{\cal B}_2\gamma_B^2},
\end{equation}
where 
\begin{equation}
\label{gammab}
\gamma_B = \frac{|e|v_g^2B}{\alpha_g^2cT^2},
\end{equation}
and, again neglecting screening effects, ${{\cal B}_1\approx 0.0037}$
and ${{\cal B}_2\approx 0.0274}$. The field dependence of the shear
viscosity at charge neutrality is illustrated in Fig.~\ref{fig:vb0}.

\subsection{Viscosity away from charge neutrality}

Away from charge neutrality, the shear viscosity (\ref{etaB0}) has to
be evaluated numerically, with the exception of the so-called
``Fermi-liquid'' or degenerate regime, ${\mu\gg T}$. In that limit,
the momentum integral in Eq.~(\ref{ttauij}) is dominated by the
momenta ${Q>W}$ [for definition of the dimensionless variables see
  Eq.~(\ref{dv})] and can be evaluated analytically. For example (here
$x=\mu/T$),
\begin{subequations}
\label{tt33fl}
\begin{equation}
\frac{1}{\tilde\tau_{33}} = \frac{16N}{\pi} \alpha_g^2\mu
\int\limits^\infty_0\frac{W^2dW}{\sinh^2W}
J_{33}(\alpha_g, W, x),
\end{equation}
where (here $\widetilde{Q}=Q/x$, $\widetilde{W}=W/x$)
\begin{equation}
J_{33}
\!=
\!\!\int\limits^1_{\widetilde{W}}\!\!
\frac{d\widetilde{Q}}{\widetilde{Q}}|\widetilde{U}|^2\!
\left(1\!-\!\frac{\widetilde{W}^2}{\widetilde{Q}^2}\right)\!
\left(1\!-\!\widetilde{Q}^2\right)^{\!2}\!
\left[1
\!-\!
\frac{\widetilde{W}^2}{\widetilde{Q}^2}\!
\left(1\!-\!\widetilde{Q}^2\right)\!\right]\!.
\end{equation}
\end{subequations}
Unscreened Coulomb interaction corresponds to $\widetilde{U}=1$ (see
below for a discussion of the screening effects).

\begin{figure}[t]
\centerline{\includegraphics[width=0.84\columnwidth]{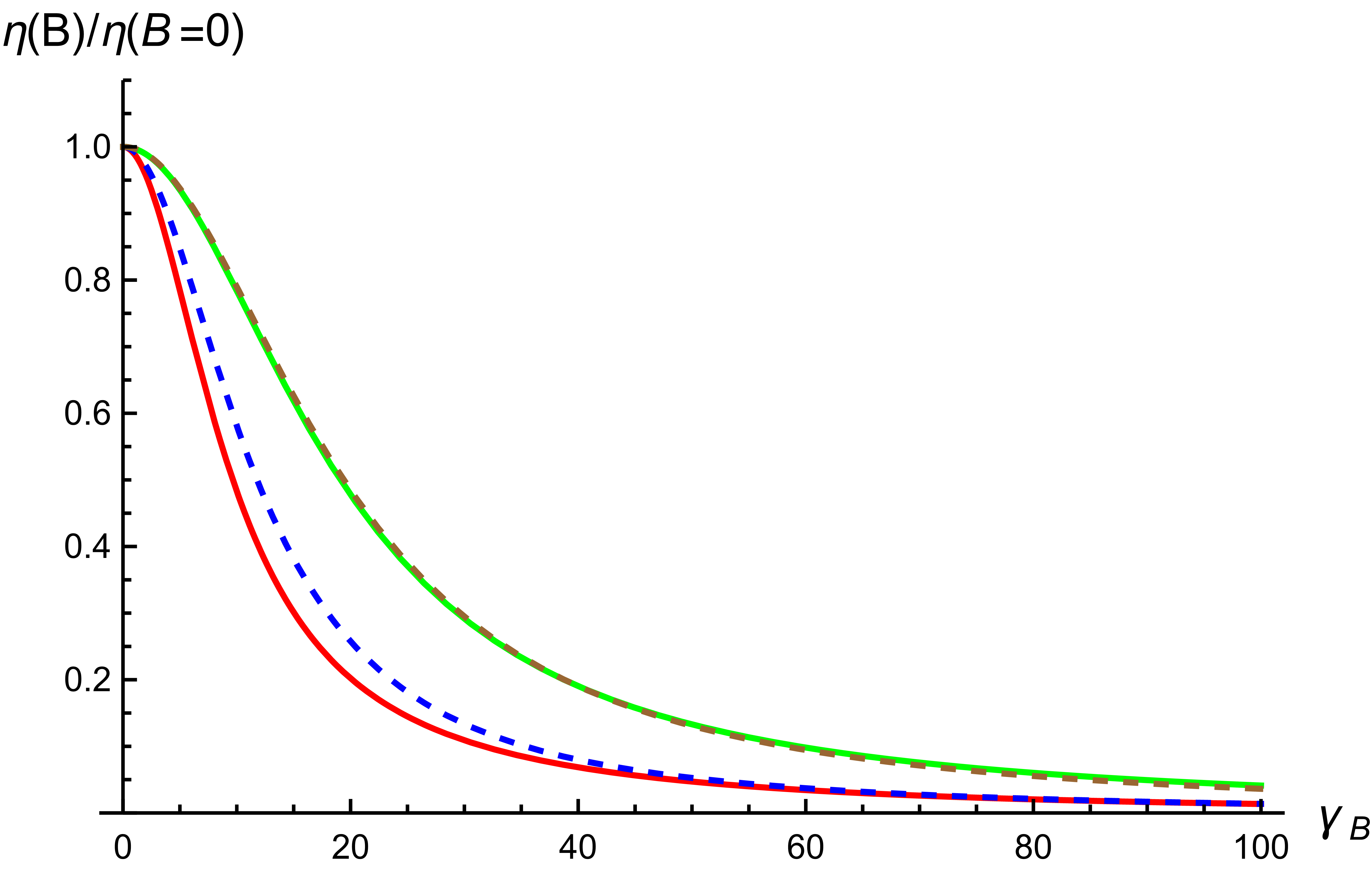}
}
\caption{Field-dependent shear viscosity in graphene away from charge
  neutrality. Solid curves are the result of the direct numerical
  evaluation of Eq.~(\ref{etaB0}) with ${\widetilde{U}=1}$. Dashed
  curves show the ``Fermi-liquid'' asymptotics (\ref{etaflB}). The
  lower (red) solid curve corresponds to ${\mu/T=3}$, the upper
  (green) -- to ${\mu/T=6}$.}
\label{fig:etaBfl}
\end{figure}

Evaluating the momentum integral $J_{33}$ for unscreened Coulomb
interaction in the limit ${\mu\gg T}$ yields
\[
J_{33}(\alpha_g\!=\!0, x\gg1)\approx \ln\frac{x}{W} - \frac{3}{2},
\]
which together with the rest of the integrals (\ref{ttauij}) leads to
the following expression for the shear viscosity
\begin{equation}
\label{vfl}
\eta(\mu\gg T) 
=
\frac{3\mu^4}{128\pi^2\alpha_g^2 v_g^2T^2}\frac{1}{\ln\frac{\mu}{T}+\delta_1-\frac{7}{4}},
\end{equation}
where ${\delta_1\approx0.34}$. The numerical factors in the
denominator represent small corrections to the leading behavior that
have to be taken into account since otherwise the matrix of the
scattering rates (\ref{taueta}) is degenerate.

Once the magnetic field is applied, the shear viscosity (\ref{etaB0})
decreases, while the Hall viscosity (\ref{etaH0}) becomes nonzero. In
classically strong fields both viscosities vanish (in contrast to the
behavior at $\mu=0$). In the degenerate regime, the field dependence
of the viscosity coefficients follows the simple semiclassical
expectation \cite{geim4,ale,moo} (similarly to the Drude conductivity
tensor):
\begin{subequations}
\label{etaflB}
\begin{equation}
\eta(B; \mu\gg T) = \frac{\eta(B=0; \mu\gg T)}{1+\Gamma_B^2},
\end{equation}
\begin{equation}
\label{etahflB}
\eta_H(B;\mu\gg T)=\eta(B=0; \mu\gg T)\frac{\Gamma_B}{1+\Gamma_B^2},
\end{equation}
where
\begin{equation}
\label{gb}
\Gamma_B
=
2 \omega_B \tilde\tau_{11},
\quad
\omega_B=|e|v_g^2B/(\mu c).
\end{equation}
\end{subequations}
The field dependence (\ref{etaflB}) was suggested in
Refs.~\onlinecite{stein,ale} for a single-component Fermi liquid. In
graphene in the degenerate limit, essentially only the single band
contributes to physical observables and hence one expects to recover
the single-band results. The kinetic calculation allows us to give a
precise definition to the scattering rate $\tilde\tau_{11}$ appearing
in Eqs.~(\ref{etaflB}). Indeed, this rate differs \cite{poli16,hydro1}
from the transport scattering rate \cite{schutt,drag}, determining the
electrical conductivity as well as from the ``quantum'' scattering
rate \cite{schutt} determining the quasiparticle lifetime.

\begin{figure}[t]
\centerline{\includegraphics[width=0.9\columnwidth]{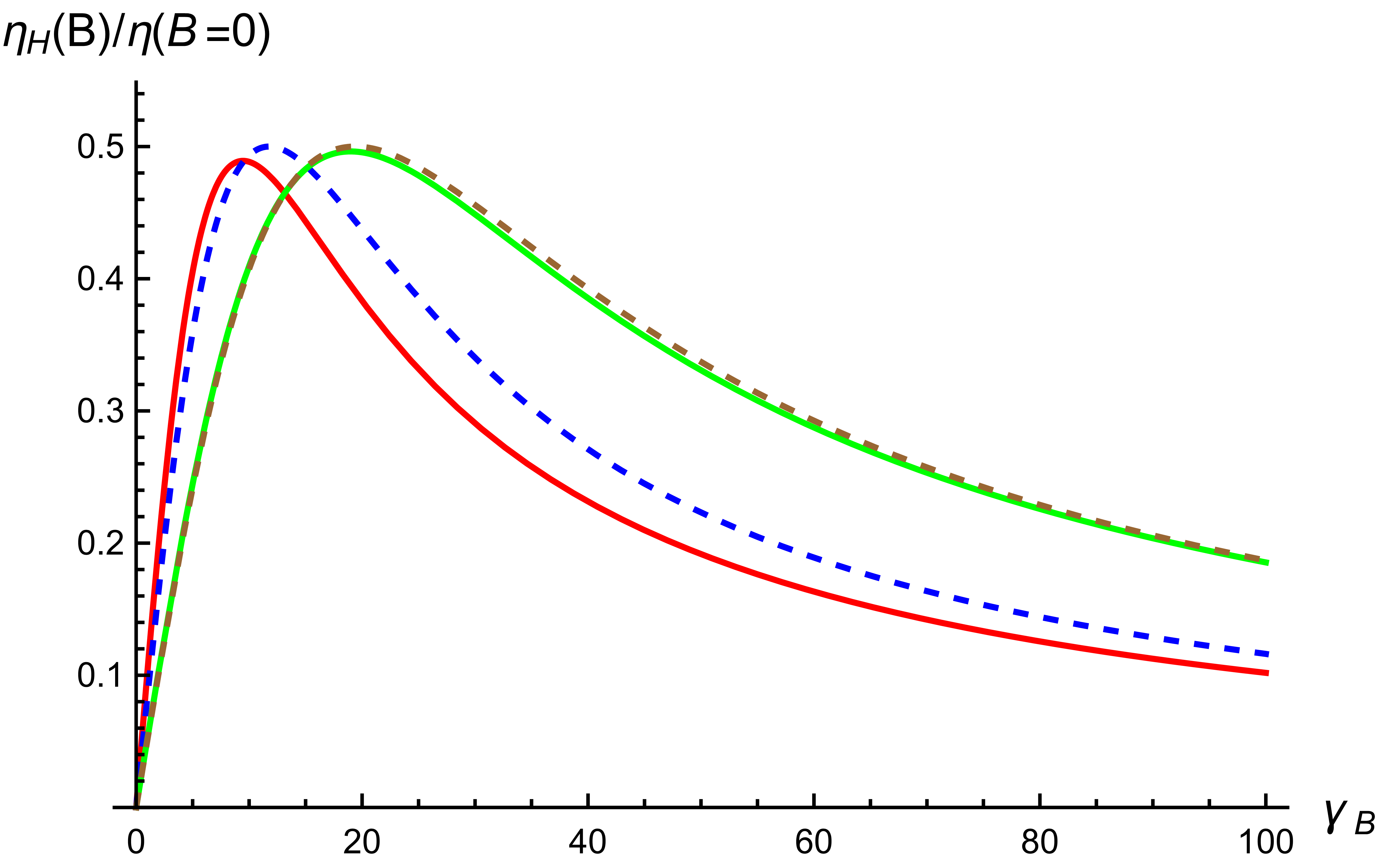}
}
\caption{Hall viscosity in graphene as a function of the magnetic
  field. Solid curves are the result of the direct numerical
  evaluation of Eq.~(\ref{etaH0}). Dashed curves show the
  ``Fermi-liquid'' asymptotics (\ref{etaflB}). The lower (red) solid
  curve corresponds to $\mu/T=3$, the upper (green) -- to $\mu/T=6$.}
\label{fig:etaHfl}
\end{figure}

In Figs.~\ref{fig:etaBfl} and \ref{fig:etaHfl} we compare the results
of the numerical evaluation (with unscreened Coulomb interaction, for
simplicity) of the shear viscosity (\ref{etaB0}) and Hall viscosity
(\ref{etaH0}) with the asymptotic expressions
(\ref{etaflB}). Qualitatively, the shape of the field dependence is
the same for all values of the chemical potential. The semiclassical
dependence (\ref{etaflB}) becomes indistinguishable from the full
result at $\mu/T\approx5$.

\section{Renormalization group approach}

Quantitative evaluation of the shear and Hall viscosities
(\ref{etaB0}) and (\ref{etaH0}) in physical units requires the
knowledge of the coupling constant $\alpha_g$. Following
Refs.~\onlinecite{shsch,msf,julia} we treat this constant as a running
coupling constant in the sense of the renormalization group (RG). The
final values for physical observables are then obtained by combining
the RG with the scaling laws for these quantities.

The one-loop Kadanoff-Wilson RG approach to interacting Dirac fermions
in graphene was suggested in
Refs.~\onlinecite{shsch,paco99,paconpb}. The idea is to relate
physical observables to their counterparts at the specifically chosen
renormalization scale, where the renormalized theory is characterized
by {\it weak coupling} and the kinetic theory is justified.

The renormalized carrier density obeys the relation
\cite{shsch}
\begin{equation}
\label{nr0}
n(T,\mu, \alpha_g) = b^{-2} n\left[ Z_T^{-1}(b) T,  Z_T^{-1}(b) \mu, \alpha_g(b)\right].
\end{equation}
Here ${T(b)=Z_T^{-1}(b)T}$ and
${\alpha_g(b)=b^{-1}Z_T^{-1}(b)\alpha_g}$ are the solutions of the RG
equations. The latter are only valid in the low-temperature quantum
limit, $T(b)<T_*$, where $T_*$ is related to the bandwidth. Choosing
the renormalization condition for ${T=0}$ and ${\mu>0}$, one finds
${\mu(b_*)=T_*}$, leading to the zero-temperature carrier density
\cite{shsch}
\[
n(T=0;\mu)=\frac{N\mu^2}{4\pi v_g^2}
\left(1\!+\!\frac{\alpha_g}{4}\ln\frac{T_*}{\mu}\right)^{\!\!-2}.
\]
For the nondegenerate system, ${\mu\ll{T}}$, the renormalization
condition is essentially the same as at criticality \cite{andy} (e.g.,
neutral graphene), ${T(b_*)=T_*}$, leading to
\[
b_*=\frac{T_*}{T}\left(1\!+\!\frac{\alpha_g}{4}\ln\frac{T_*}{T}\right)\!,
\quad
Z_T(b_*) = \frac{1}{b_*}\left(1\!+\!\frac{\alpha_g}{4}\ln\frac{T_*}{T}\right)\!.
\]
For general $\mu$ and $T$, the leading behavior is captured by
\begin{equation}
\label{nr1}
n(T,\mu) \!=\! 
\frac{NT^2}{2\pi v_g^2} \frac{\tilde{n}(x)}{R_\Lambda^2},
\quad
R_\Lambda\!=\!1\!+\!\frac{\alpha_g}{4}\ln\frac{T_*}{{\rm max}(\mu,T)}\!,
\end{equation}
where $\tilde{n}(x)$ is a dimensionless function of ratio ${x=\mu/T}$
that can be read off Eq.~(\ref{n0}) (for ${\bs{u}=0}$). The ratio
${x=\mu/T}$ does not change under the RG, hence only the explicit
dependence on the velocity $v_g$ is renormalized \cite{kash}.
At the same time, the renormalized coupling constant is
\begin{equation}
\label{ar}
\alpha_g = \alpha_g^{(0)}/R_\Lambda.
\end{equation}

The ``bare'' coupling constant in suspended graphene is ${e^2/(\hbar
  v_g)\approx2.2}$ (corresponding to the ``bare'' velocity
${v_g\approx10^6}\,$m/s), while for graphene encapsulated in boron
nitride \cite{geim1} this reduces to ${e^2/(\hbar \varepsilon
  v_g)\approx0.5}$ (here $\varepsilon$ is the effective dielectric
constant). Assuming the value ${T_*\approx8.34\!\times\!10^4\,}$K, see
Refs.~\onlinecite{shsch,paco}, and ${\alpha_g^{(0)}\approx0.5}$ we
estimate the effective coupling constant in neutral graphene
\[
\alpha_g(T\approx200\,{\rm K}) \approx 0.285,
\qquad
\alpha_g(T\approx10\,{\rm mK}) \approx 0.167.
\]
Even assuming a smaller, ``Fermi-liquid'' coupling constant,
${\alpha_g^{(0)}\approx0.2}$, derived from earlier measurements
\cite{sav,meg}, we find the renormalized value of $0.11$ at the lowest
temperatures and $0.15$ in the hydrodynamic range.

\begin{figure}[t]
\centerline{\includegraphics[width=0.92\columnwidth]{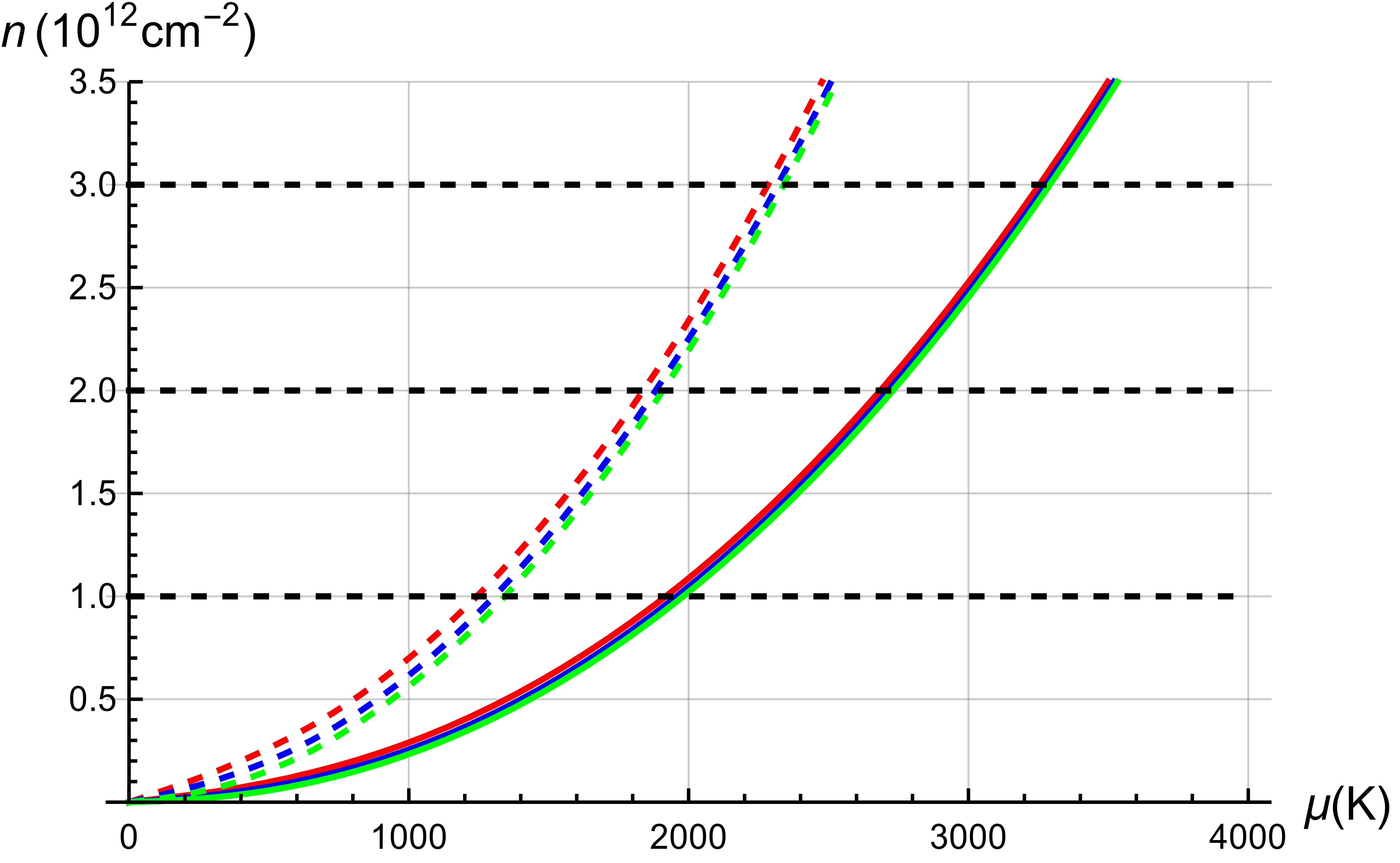}
}
\caption{Renormalized carrier density (\ref{nr1}). The red, blue, and
  green solid curves correspond to ${T=300,200,100}\,$K,
  respectively. Dashed curves show the nonrenormalized density.}
\label{fig:ntr}
\end{figure}

For realistic values of the carrier density in graphene in the
degenerate regime \cite{geim1,geim2,geim3,geim4}, the logarithmic
renormalization (\ref{nr1}) is appreciable, see
Fig.~\ref{fig:ntr}. For ${T=300}\,$K, the dimensionful prefactor in
Eq.~(\ref{nr1}) has the value
\[
\frac{NT^2}{2\pi v_g^2} 
\rightarrow
\left.\frac{2k_B^2T^2}{\pi\hbar^2 v_g^2} \right|_{T=300\,{\rm K}}
\approx
0.098\!\times\! 10^{12} \, {\rm cm}^{-2}.
\]
Neglecting the renormalization factor in Eq.~(\ref{nr1}) and using the
``Fermi-liquid'' asymptotics, ${\tilde{n}=x^2/2+\pi^2/3}$, we estimate
the chemical potential corresponding to the typical density,
${n=10^{12}\,}$cm$^{-2}$, as
\[
\mu = \sqrt{\pi k_B^{-2}\hbar^2v_g^2n\!\left(\!1\!-\!\frac{\pi k_B^2T^2}{3\hbar^2v_g^2n}\right)}
\approx 1239.7 \, {\rm K},
\quad
x \approx 4.13.
\]
Restoring the renormalization factor, we find increased values for the
chemical potential, see Fig.~\ref{fig:ntr} (the values shown in the
figure were evaluated for ${\alpha_g^{(0)}\approx0.5}$). For the same
carrier density we find
\[
\mu\left(n\!=\!10^{12}{\rm cm}^{-2}\right)\approx1918.8\,{\rm K},
\quad
x\approx6.396,
\quad
R_\Lambda
\approx
1.47,
\]
with the corresponding renormalized coupling constant
\[
\alpha_g \approx 0.34.
\]
For the smaller ``bare'' coupling constant,
${\alpha_g^{(0)}\approx0.2}$, the above values change to
\[
\mu\left(n\!=\!10^{12}{\rm cm}^{-2}\right)\approx1533.1\,{\rm K},
\quad
\alpha_g \approx 0.167,
\quad
R_\Lambda
\approx
1.2.
\]
Neither renormalization factor is negligible leading to a strong
enhancement of the results of the kinetic theory.

Similarly to Eq.~(\ref{nr0}), the viscosity renormalizes as
\cite{msf,julia}
\begin{equation}
\label{etar0}
\!
\eta(T,\mu,\alpha_g, B) \!=\! \frac{1}{b^2}
\eta\left[ Z_T^{-1}(b) T,  Z_T^{-1}(b) \mu, \alpha_g(b), b^2B\right]\!,
\end{equation}
where $B$ is the magnetic field. Representing the weak-coupling
kinetic-theory result (\ref{etaB0}) as
\begin{equation}
\label{etar1}
\eta = \frac{T^2}{v_g^2\alpha_g^2} \tilde\eta(x,\Gamma_B^{ij}),
\end{equation}
one finds \cite{kash,msf} that the product $v_g\alpha_g$ doesn't
renormalize. This leads to the conclusion \cite{shsch} that the
kinetic expression for the shear viscosity at charge neutrality,
Eq.~(\ref{vdpb}), provides the correct low temperature result for
interacting Dirac fermions in graphene (for ${B=0}$). Implicit to this
argument is the assumption of the large screening length \cite{kash}.
Indeed, static screening is determined by the real part of the
polarization operator, ${\Pi^R(\omega=0)}$, which for small enough
momenta is well approximated by the density of states \cite{drag},
\[
\varkappa = 2\pi\alpha_gv_g\Pi^R(\omega\!=\!0)\approx
4\alpha_g\frac{\cal T}{v_gR_\Lambda^2},
\]
where ${{\cal T}=2T\ln[2\cosh(\mu/2T)]}$, see Eq.~(\ref{t}). The
scaling of the density of states was derived in
Refs.~\onlinecite{shsch,hwahu}.

The quantities $\Gamma_B^{ij}$ are defined similarly to Eq.~(\ref{gb})
\begin{equation}
\label{gbr}
\Gamma_B^{ij} = 2 \omega_B \tilde\tau_{ij} = \frac{|e|v_g^2B}{c\alpha_g^2T^2} 
f_{ij}(x, \varkappa)R_\Lambda^{4},
\end{equation}
where the dimensionless functions $f_{ij}(x, \varkappa)$ can be read off
Eqs.~(\ref{gb}) and (\ref{ttauij}).

Finally, the kinematic viscosity (\ref{kv}) is a combination of the
shear viscosity, renormalized velocity, and energy density. The latter
renormalizes as the free energy \cite{shsch}
\begin{equation}
\label{nEr0}
n_E(T,\mu, \alpha)
\!=\!
\frac{Z_T(b)}{b^2} n_E\left[ Z_T^{-1}(b) T,  Z_T^{-1}(b) \mu, \alpha_g(b)\right]\!,
\end{equation}
which yields
\begin{equation}
\label{nEr1}
n_E(T,\mu, \alpha)
\!=\! 
\frac{NT^3}{\pi v_g^2} \frac{\tilde{n}_E(x)}{R_\Lambda^2}.
\end{equation}
The kinematic viscosity (\ref{kv}) can be obtained by combining
Eqs.~(\ref{etar1}) and (\ref{nEr1}) with the equation of state
(\ref{eqsta}). The result is given by
\begin{equation}
\label{nur1}
\nu(T,\mu, \alpha)
\!=\!
\frac{2\pi v_g^2\tilde\eta(x,\Gamma_B^{ij})}{3N\alpha_g^2T\tilde{n}_E(x)} R_\Lambda^{4}.
\end{equation}

The results of the one-loop RG approach reviewed in this section
should be treated with care. Ultimately, this is a perturbative
calculation that is formally valid for weak coupling. Strictly
speaking, this is not the case in real graphene (see the above
estimates for the effectiv coupling constant) and hence the kinetic
coefficients, such as viscosity, should be considered
phenomenologically and assigned the experimentally measured values
\cite{luc}. Nevertheless, it is instructive to evaluate the
expressions for kinetic coefficients with the corresponding
renormalizations to obtain a quantitative theoretical expectation for
physical observables.

\section{Quantitative results for electronic viscosity in graphene}

In this section we report the results of the numerical evaluation of
the shear and Hall viscosity in graphene (\ref{etares}) taking into
account the renormalization and screening effects.

\subsection{Screening effects}

The above analytical results were obtained for the bare Coulomb
interaction and are valid only in the limit of the infinitesimal
interaction strength, $\alpha_g\rightarrow0$. For more realistic
values of $\alpha_g$, screening effects have to be taken into
account. Within the RPA approximation \cite{schutt,drag,luc} the
dynamically screened Coulomb interaction is given by
\begin{equation}
\label{rpa}
{\cal D}^R_{RPA}(\omega, \bs{q}) = \frac{U_0}{1\!+\!U_0\Pi^R(\omega, \bs{q})},
\qquad
U_0=\frac{2\pi e^2}{q},
\end{equation}
where $\Pi^R(\omega, \bs{q})$ is the polarization operator (for a
detailed calculation of the polarization operator see, e.g.,
Ref.~\onlinecite{drag}). In the dimensionless form of
Eq.~(\ref{tt33fl}) one finds
\[
\widetilde{U}=\frac{Q}{Q\!+\!2\pi\alpha_g\widetilde\Pi^R(\omega, \bs{q})},
\qquad
\widetilde\Pi^R(\omega, \bs{q}) = \frac{v_g^2}{2T}\Pi^R(\omega, \bs{q}).
\]

For analytical estimates in the degenerate regime one can use the
usual Thomas-Fermi static screening \cite{schutt,drag}
\[
\widetilde{U}=\frac{\widetilde{Q}}{\widetilde{Q}\!+\!2\alpha_g}.
\]
In full units the inverse screening length is
${\varkappa=N\alpha_gk_F}$. The use of the static screening in
Eq.~(\ref{tt33fl}) can be justified by the fact that in this integral
the contribution of the region ${Q\sim W}$ is explicitly suppressed,
while outside of this region for ${W<Q<x}$ the polarization operator
in graphene is well approximated by a constant \cite{schutt,drag}.

Taking into account static screening, we find for the momentum
integral $J_{33}$ in the opposite order of limits (first,
${x\rightarrow\infty}$, then ${\alpha\to0}$)
\[
\lim_{\alpha_g\rightarrow0}\lim_{x\rightarrow\infty} J_{33}
\approx
\ln\frac{1}{2\alpha_g} - \frac{7}{4}
+ \frac{W^2}{2x^2\alpha_g^2}\left(1-\ln\frac{2\alpha_gx}{W}\right).
\]
The first two terms are identical for all integrals (\ref{taueta}), so
again we need to keep the subleading term. Combined with the rest of
the integrals (\ref{ttauij}) this leads to the result
\begin{equation}
\label{vfls}
\eta(\mu\gg T) 
=
\frac{3\mu^4}{128\pi^2\alpha_g^2 v_g^2T^2}\frac{1}{\ln\frac{1}{\alpha_g}-\delta_2},
\end{equation}
where ${\delta_2=7/4+\ln2\approx2.44}$. The factor $\delta_2$ is kept
in the denominator in order to emphasize that the logarithmic
dependence on the coupling constant is only valid in the limit
${\alpha_g\rightarrow0}$ such that ${|\ln\alpha_g|\gg1}$. For any
practical value of the coupling constant Eq.~(\ref{vfls}) is negative
and thus invalid. Instead of the limit ${\alpha_g\rightarrow0}$, one
has to consider the full expression for $J_{33}$.

The leading parametric dependence in Eq.~(\ref{vfls}) (up to the
correction, $\delta_2$) was first suggested in
Ref.~\onlinecite{poli16}. The numerical prefactor that can be found in
Ref.~\onlinecite{luc} appears to be twice sa large as ours.

\begin{figure}[t]
\centerline{\includegraphics[width=0.95\columnwidth]{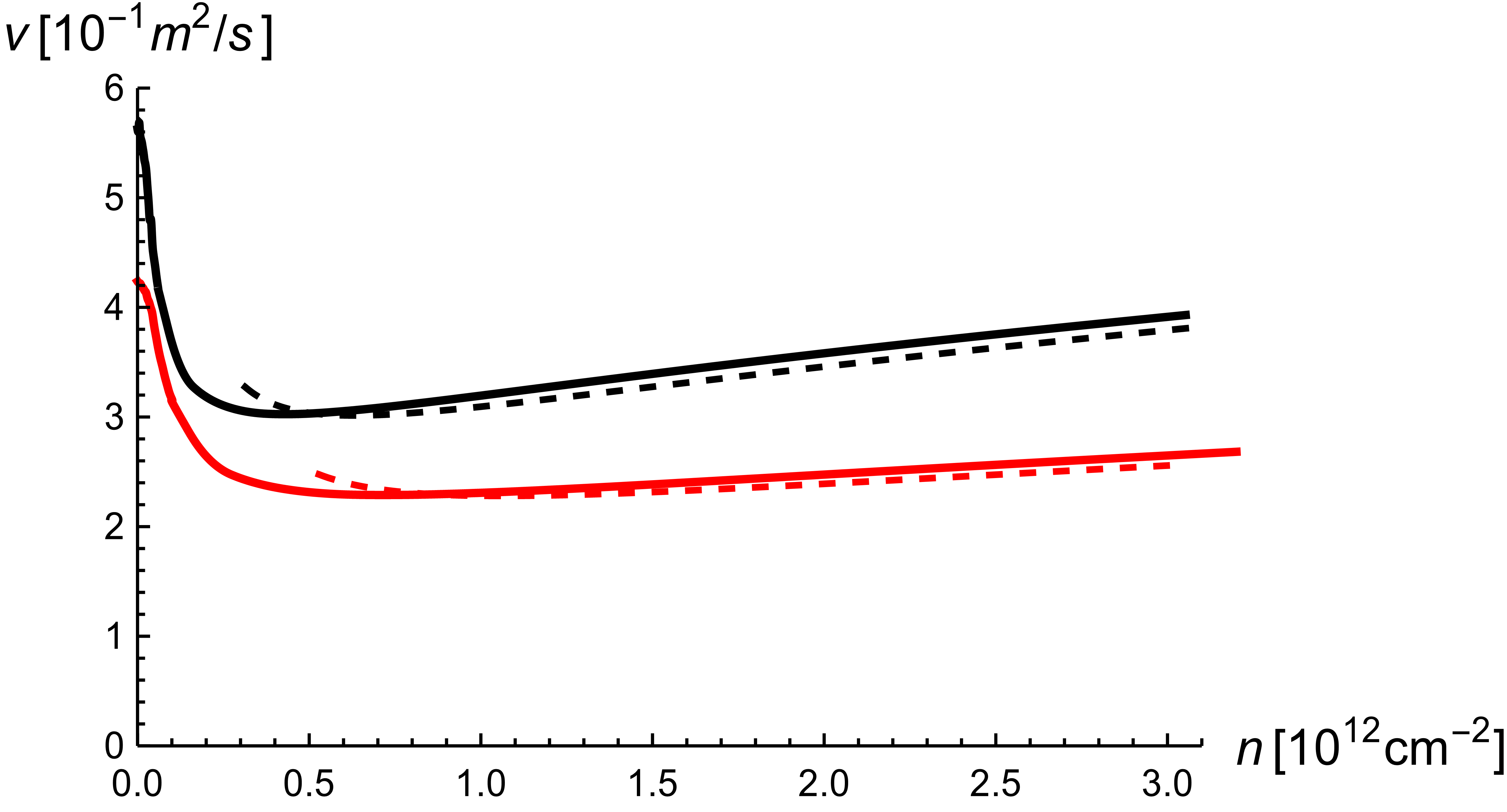}
}
\caption{Kinematic viscosity in graphene as a function of the carrier
  density. The solid lines are the result of the direct numerical
  evaluation of Eq.~(\ref{nur1}) with ${\alpha_g=0.5}$ and dynamical
  screening, Eq.~(\ref{rpa}). Dashed curves show the ``Fermi-liquid''
  asymptotics, based on Eq.~(\ref{tt33fl}) with static screening. The
  lower dataset (shown in red) corresponds to ${T=280}\,$K, the upper
  (black) -- to ${T=220}\,$K. The temperatures and the density range
  are taken from Fig.~4 of Ref.~\onlinecite{geim1}.}
\label{fig:nux}
\end{figure}

In gated structures screening is modified by the presence of the gate
\cite{ash}. In particular, the ``bare'' Coulomb interaction $U_0$
should be replaced by
\begin{equation}
\label{gc}
U_0\rightarrow U_0^g = \frac{2\pi e^2}{q} \left(1-e^{-2qd}\right),
\end{equation}
where $d$ is the distance to the gate. This form assumes the single
gate device. Note, that the experiments of Ref.~\onlinecite{geim1}
were performed on double gate devices as well. In the latter case, the
effective Coulomb interaction has a more complicated form. However, if
the gate is placed far enough from the graphene sample
(Ref.~\onlinecite{geim1} reports the thickness of the insulating layer
to be about $d=50\,$nm), such that $d\gg1/\varkappa$, then the
screening effect of the gate may be neglected.

\subsection{Kinematic viscosity in zero field}

Kinematic viscosity in graphene as a function of the carrier density
is shown in Fig.~\ref{fig:nux}, where we plot our results in physical
units taking into account the renormalizations (\ref{nr1}) and
(\ref{nur1}) as well as the dynamical screening (\ref{rpa}). The
results are in a reasonably good agreement with the experimental data
reported in Fig. 4 of Ref.~\onlinecite{geim1}: the theoretical values are
of the same order of magnitude, ${\nu\sim0.1}\,$m$^2$/s, as the
experimental ones, the density dependence in the range
${n\approx1\div3\times10^{12}}\,$cm$^{-2}$ is rather weak, and the
overall value decreases slightly with the temperature increase from
${T=220}\,$K to ${T=280}\,$K.

Our results were obtained assuming the value ${\alpha_g\approx0.5}$
for the ``bare'' coupling constant (as reported in the Supplemental
Material to Ref.~\onlinecite{geim1}). The ``bare'' velocity in
graphene was taken as $v_g=10^6\,$m/s. Then at ${T=300}\,$K, the
dimensionful prefactor in Eq.~(\ref{nur1}) has a value
\[
\frac{v_g^2}{\alpha_g^2T} \rightarrow
\left.\frac{\hbar v_g^2}{\alpha_g^2k_BT}\right|_{T=300\,{\rm K}}
\approx
0.1 \frac{{\rm m}^2}{\rm s},
\]
which ultimately determines the order of magnitude of the resulting
kinematic viscosity.

\begin{figure}[t]
\centerline{\includegraphics[width=0.9\columnwidth]{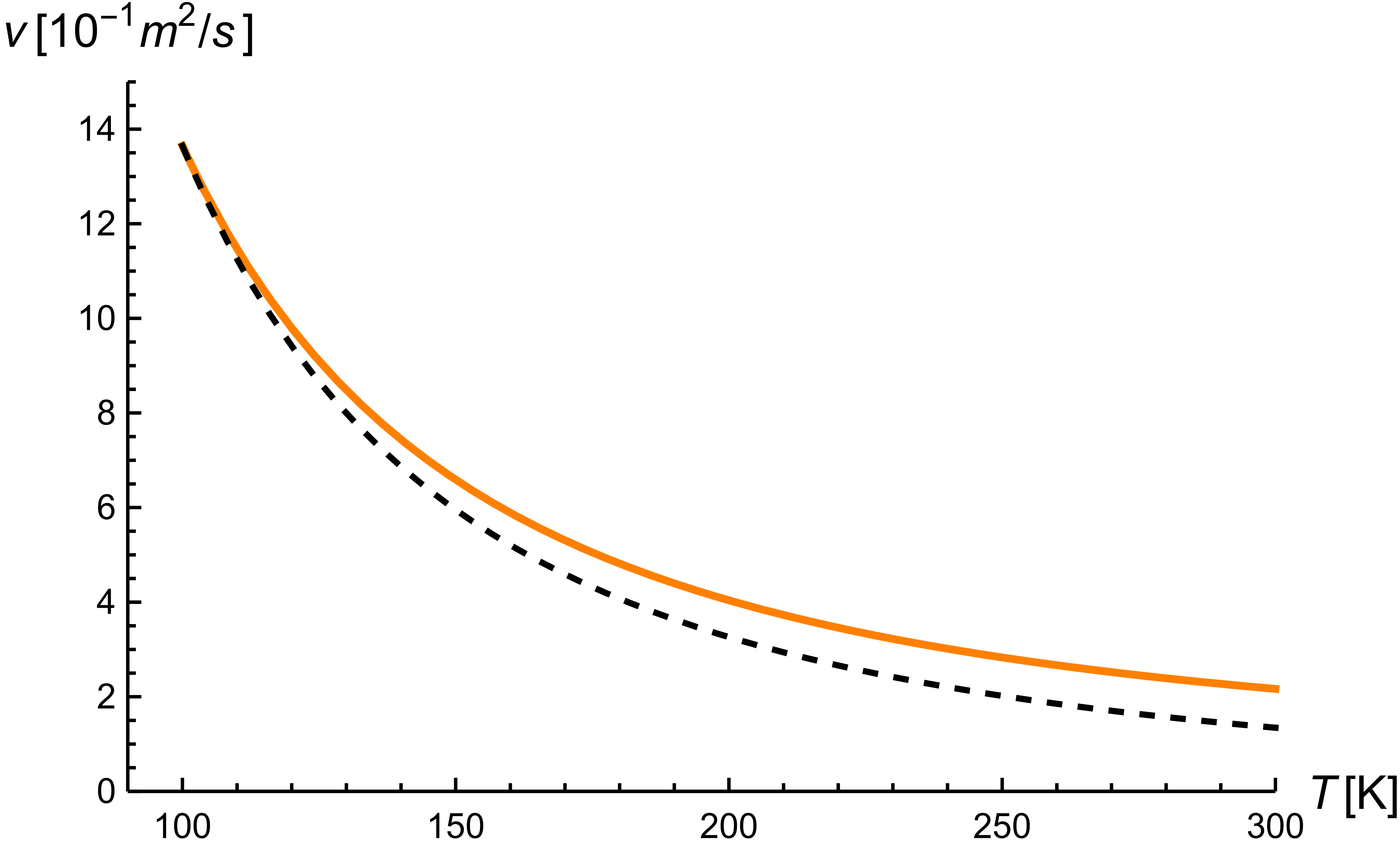}
}
\caption{Kinematic viscosity in graphene as a function of temperature
  evaluated at the carrier density ${n=2\times10^{12}}\,$cm$^{-2}$,
  where the ``Fermi-liquid'' asymptotics yields excellent agreement
  with the direct numerical evaluation of Eq.~(\ref{nur1}). The solid
  line shows the ``Fermi-liquid'' asymptotics for the kinematic
  viscosity (\ref{kv}) based on Eq.~(\ref{tt33fl}). The dashed line
  shows the ``naive'' temperature dependence (\ref{ntd}), vertically
  shifted for clarity. The temperature and density ranges are taken
  from Fig. 3 of Ref.~\onlinecite{geim4}.}
\label{fig:etat}
\end{figure}

Combining this prefactor with the asymptotic behavior of $\tilde\eta$
[which can be read off Eqs.~(\ref{vfl}) or (\ref{vfls}) disregarding
  the logarithm] and ${\tilde n_E\approx{x}^3(1\!+\!\pi^2/x^2)/6}$
[from Eq.~(\ref{ne0})], we estimate the dominant temperature
dependence of the kinematic viscosity (\ref{kv}) in the degenerate
regime as
\begin{equation}
\label{ntd}
\nu(\mu\gg T)\propto \frac{v_g^2\mu}{\alpha_g^2T^2}\frac{1}{1+\pi^2T^2/\mu^2}.
\end{equation}
This ``naive'' estimate neglects the temperature dependence arising
from the renormalizations and the logarithmic factors in
Eq.~(\ref{vfl}) [or Eq.~(\ref{vfls})]. Nevertheless, the true
temperature dependence is not far off as shown in Fig.~\ref{fig:etat}
where we plot Eq.~(\ref{ntd}) together with the ``Fermi-liquid''
asymptotics for the kinematic viscosity (\ref{kv}) based on
Eq.~(\ref{tt33fl}) and static screening (a reasonable approximation in
the degenerate regime, see Fig.~\ref{fig:nux}; the dashed curve is
vertically shifted for clarity).

Similarly, we can estimate the temperature dependence of the kinematic
viscosity at charge neutrality. Using Eq.~(\ref{vdp}) and the
relation ${n_E(\mu\!=\!0)\propto T^3}$, we find
\begin{equation}
\label{ntddp}
\nu(\mu=0) \propto \frac{v_g^2}{\alpha_g^2T}.
\end{equation}
Again, the true temperature dependence will be slightly different due
to the renormalization factors (the screening length at charge
neutrality is determined by temperature and hence does not lead to any
additional temperature dependence).

For the higher temperature range shown in Fig.~\ref{fig:etat} the
kinematic viscosity may be fitted \cite{poli16} by another power law,
${\nu\propto{T}^{-1.5}}$. However, this is an intermediate regime: for
higher temperatures, $T\geqslant350\,$K, the calculated temperature
dependence shows clear deviations from this behavior.

Finally, typical theoretical values of the kinematic viscosity shown
in Fig.~\ref{fig:nux} differ from those reported in
Ref.~\onlinecite{geim1} by about a factor of $3$. Our calculation does
not involve any fitting parameters and does not take into account any
particular features of the experimental device, e.g., screening by the
gate and disorder scattering. As we have already mentioned, the
renormalization group calculation leading to the factors of
$R_\Lambda$ is approximate, so that we do not expect to find a perfect
agreement with the data. A true test of the theory would be to
calculate the quantities actually measured in the experiment (e.g.,
the nonlocal resistivity \cite{geim1}, $R_V$) for realistic sample
geometries. The results of such calculations will be reported in a
subsequent publication.

\subsection{Hall viscosity}

Defining the ``kinematic'' counterpart of the Hall viscosity similarly
to Eq.~(\ref{kv}),
\begin{equation}
\label{kvH}
\nu_H = \frac{v_g^2\eta_H}{W},
\end{equation}
and using the same RG approach, we find that $\nu_H$ renormalizes
similarly to Eq.~(\ref{nur1})
\begin{equation}
\label{nuHr1}
\nu_H(T,\mu, \alpha)
=
\frac{2\pi v_g^2\tilde\eta_H(x,\gamma_B)}{3N\alpha_g^2T\tilde{n}_E(x)} 
R_\Lambda^4.
\end{equation}
Here $\tilde\eta_H(x,\gamma_B)$ is defined according to
Eq.~(\ref{etar1}). For ${T=300\,}$K, ${B=1\,}$T and neglecting
renormalizations, the dimensionless quantity $\gamma_B$ (in SI units)
is given by
\[
\gamma_B=
\left.\frac{\hbar|e|v_g^2B}{\alpha_g^2k_B^2T^2}\right|_{T=300\,{\rm K}; B=1\,{\rm T}}
\approx3.939.
\]

\begin{figure}[t]
\centerline{\includegraphics[width=0.915\columnwidth]{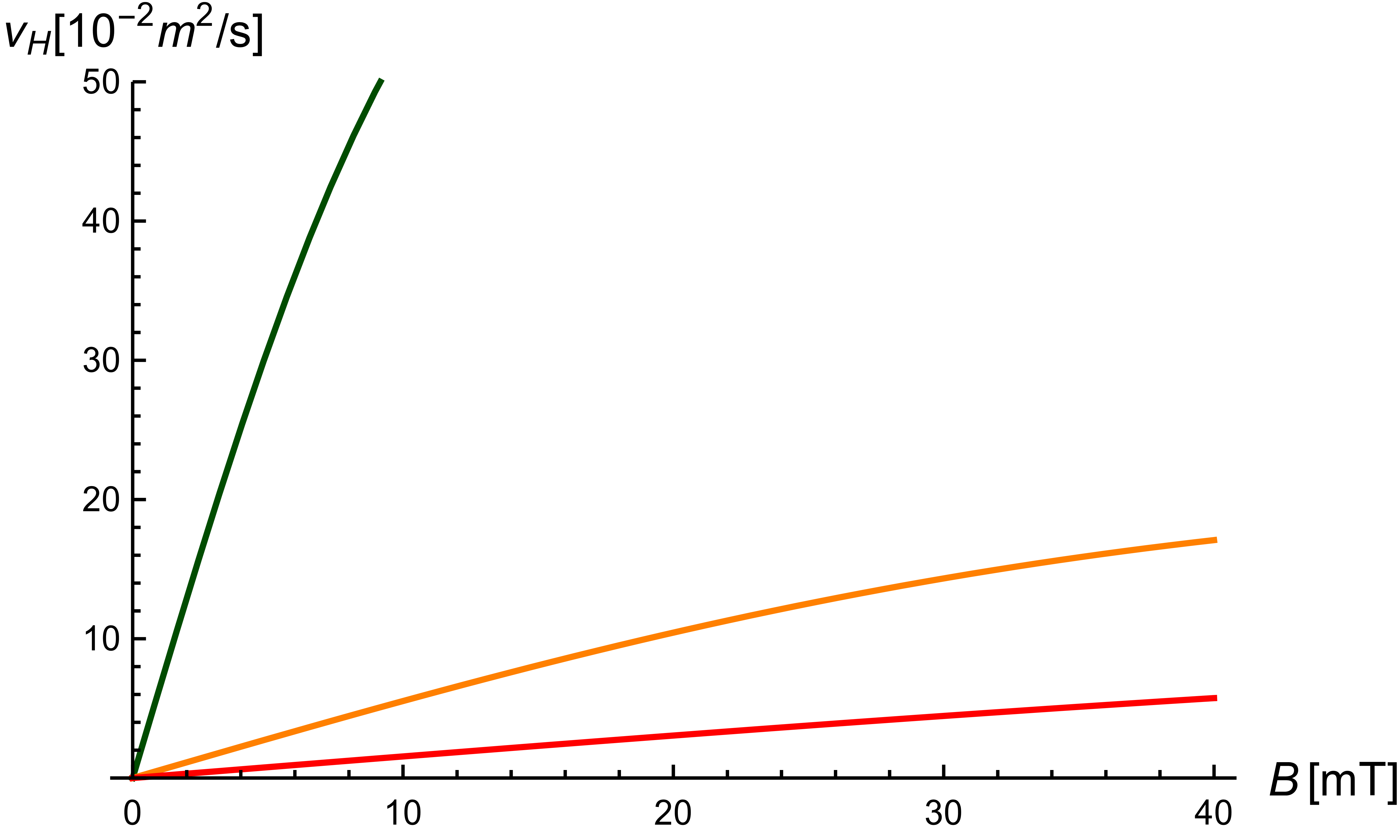}
}
\caption{Hall viscosity in graphene as a function of magnetic field.
  Solid lines show the ``Fermi-liquid'' asymptotics for the kinematic
  Hall viscosity (\ref{kvH}) based on Eq.~(\ref{tt33fl}) with
  $\alpha_g=0.5$ for ${n=2\times10^{12}}\,$cm$^{-2}$ and three
  different temperatures, ${T=100\,}$K, ${T=220\,}$K, and ${T=300\,}$K
  (shown in green, orange, and red, respectively). The density,
  temperature, and field range are taken from Fig.~3 of
  Ref.~\onlinecite{geim4}.}
\label{fig:vhb}
\end{figure}

\begin{figure}[t]
\centerline{\includegraphics[width=0.9\columnwidth]{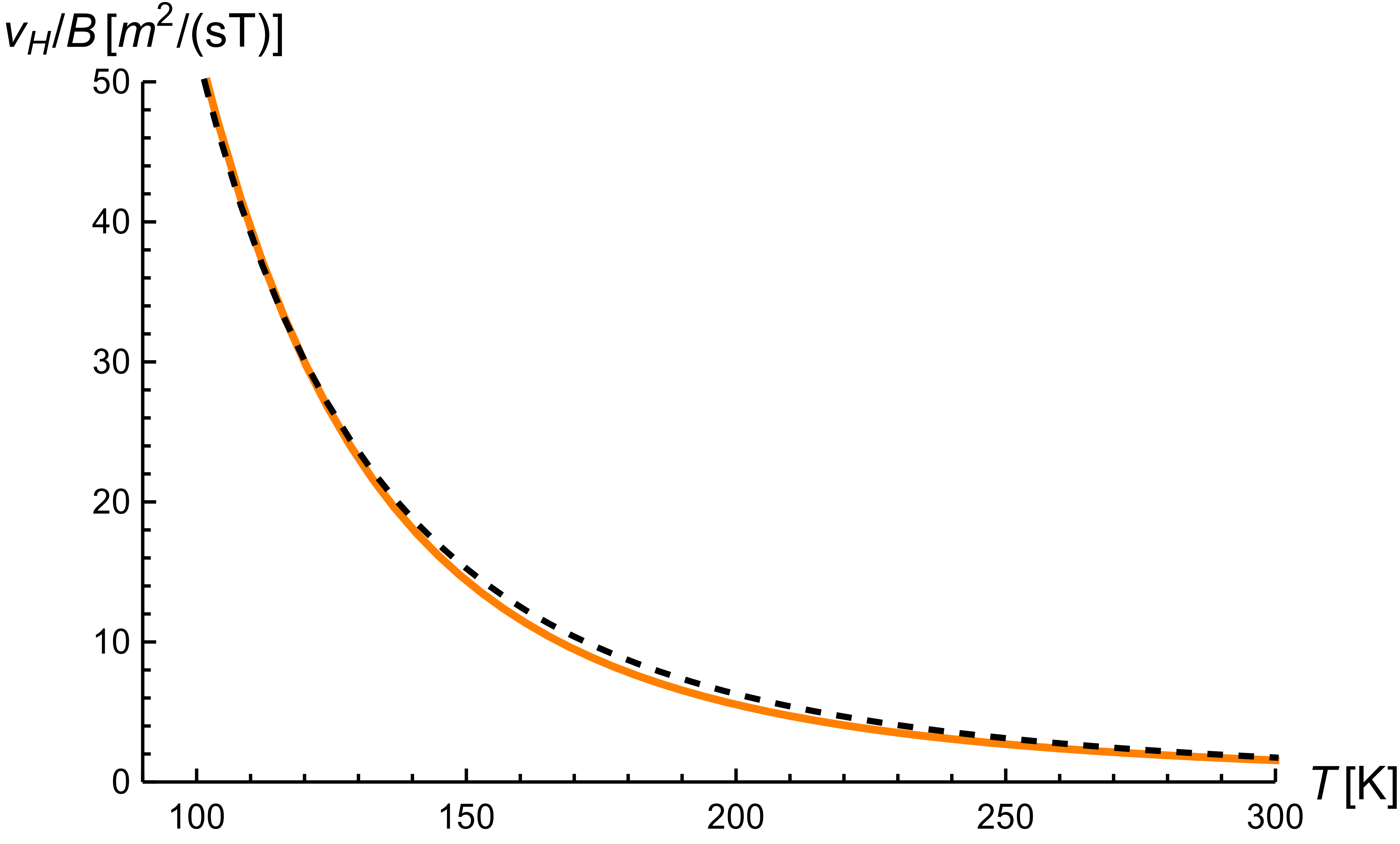}
}
\caption{Temperature dependence of the kinematic Hall viscosity. The
  solid line shows the ``Fermi-liquid'' asymptotics for the kinematic
  Hall viscosity (\ref{kvH}) based on Eq.~(\ref{tt33fl}) with
  $\alpha_g=0.5$ for ${n=2\times10^{12}}\,$cm$^{-2}$. The dashed line
  shows the ``naive'' temperature dependence (\ref{ntdH}), vertically
  shifted for clarity. The density and temperature ranges are taken
  from Fig.~3 of Ref.~\onlinecite{geim4}.}
\label{fig:vht}
\end{figure}

The resulting values are shown in Figs.~\ref{fig:vhb} and
\ref{fig:vht} in physical units. The former shows $\nu_H$ as a
function of the external magnetic field for a fixed value of the
charge density, ${n\approx2\times10^{12}}\,$cm$^{-2}$. The results
shown in Fig.~\ref{fig:vhb} significantly exceed the experimental
values shown in Fig.~3 of Ref.~\onlinecite{geim4}. The origin of this
discrepancy is in the high power of the renormalization factor in
Eqs.~(\ref{nuHr1}) and (\ref{gbr}). Indeed, combining
Eqs.~(\ref{etahflB}), (\ref{nuHr1}), and (\ref{gbr}), we find
\begin{subequations}
\label{nH}
\begin{equation}
\nu_H = R_\Lambda^4 \frac{\pi}{6} \frac{v_g^2}{\alpha_g^2T} 
\frac{\tilde\eta(x,\varkappa)}{\tilde{n}_E(x)} \frac{B B_0}{B^2\!+\!B_0^2},
\end{equation}
where the ``reference field'' $B_0$ introduced in
Refs.~\onlinecite{pol17,geim4} is given by
\begin{equation}
\label{B0}
B_0 = R_\Lambda^{-4} \frac{c\alpha_g^2T^2}{|e|v_g^2f(x,\varkappa)}.
\end{equation}
\end{subequations}
Here the dimensionless function $f(x,\varkappa)$ is defined similarly
to Eq.~(\ref{gbr}) with the dominant contribution coming from the
scattering time $\tilde\tau_{11}$ [defined in Eq.~(\ref{ttauij})].

The temperature dependence of $\nu_H$ shown in Fig.~\ref{fig:vht} for
weak magnetic fields, ${B\ll{B}_0}$, and in the degenerate regime can
be extracted from Eqs.~(\ref{nH}) and (\ref{ttauij}),
\begin{equation}
\label{ntdH}
\nu_H(\mu\gg T; B\ll B_0)\propto \frac{|e|v_g^4}{c\alpha_g^4T^3}\frac{1}{1+\pi^2T^2/\mu^2}.
\end{equation}
Similarly to Eqs.~(\ref{ntd}) and (\ref{ntddp}) this expression
disregards additional temperature dependence from the renormalizations
and screening [i.e., the logarithmic factors in Eqs.~(\ref{vfl}) or
  Eq.~(\ref{vfls})]. Nevertheless, it accounts for the temperature
dependence of the Hall viscosity in the degenerate regime rather well,
see Fig.~\ref{fig:vht}.

\section{Summary}

In conclusion, we have calculated the shear and Hall viscosities in
graphene at arbitrary doping levels and classical (nonquantizing)
magnetic fields using the kinetic theory approach combined with the RG
analysis. The shear viscosity in graphene exhibits a monotonous growth
as a function of carrier density (or chemical potential) from a small
value at charge neutrality, see Eqs.~(\ref{vfl}) and (\ref{vfls}). In
contrast, the kinematic viscosity (\ref{kv}) remains of the same order
of magnitude at all doping levels, see Fig.~\ref{fig:nux}: $\nu_H(n)$
decays from the initial value at ${n=0}$ and then passes through a
minimum followed by a (initially weak) growth in the degenerate
regime. This behavior follows from the nontrivial density dependence
of the enthalpy (or equivalently, energy density). The appearance of
the enthalpy in the definition of kinematic viscosity (\ref{kv}) is a
characteristic feature of the electronic system in graphene (in
contrast, the usual definition of the kinematic viscosity involves the
mass density \cite{dau6}).

The field dependence of the shear and Hall viscosities shown in
Figs.~\ref{fig:etaBfl} and \ref{fig:etaHfl} can be expected on general
grounds. In particular, the Hall viscosity vanishes at zero magnetic
field as well as for classically strong fields and hence has to
exhibit a maximum. In the degenerate regime, both viscosities are well
described by the semiclassical expressions (\ref{etaflB}) first
suggested in Ref.~\onlinecite{ale}. For smaller densities the shape of
the field dependence of $\eta$ and $\eta_H$ deviates from
Eqs.~(\ref{etaflB}), but remains very similar. The only exception to
this argument is the field dependence of the shear viscosity at charge
neutrality, see Fig.~\ref{fig:vb0}, which remains finite in classically
strong fields. This effect can be traced to the complete decoupling of
the charge and energy currents at charge neutrality.

In the degenerate regime of large charge densities, our results are in
a reasonably good agreement with the available experimental
evidence. The quantitative discrepancies between the theoretical
values shown in Figs.~\ref{fig:nux}-\ref{fig:vht} and the results of
Refs.~\onlinecite{geim1,geim4} can be attributed to our use of the
renormalization group resulting in relatively strong enhancement of
the kinematic viscosity (\ref{kv}) and especially the kinematic Hall
viscosity (\ref{kvH}). Moreover, our calculation does not include
sample-specific details such as screening by the gate and disorder
scattering. The latter can be expected to reduce the viscosity values.
In any case, a true test of the theory would be a calculation of
several distinct quantities actually measured in the experiment for
realistic sample geometries (using, e.g., the experimentally measured
\cite{vlan1,vgeim1,vlan2} values for renormalized velocity). Such 
results will be reported elsewhere.

\acknowledgments

We thank I.V. Gornyi, A.D. Mirlin, and J. Schmalian for fruitful
discussions. This work was supported by the German Research Foundation
DFG within FLAG-ERA Joint Transnational Call (Project GRANSPORT) and
the MEPhI Academic Excellence Project, Contract No. 02.a03.21.0005
(BNN) and the European Union's Horizon 2020 research and innovation
programme under the Marie Sklodowska-Curie Grant Agreement No. 701647
(MS).

\appendix

\section{Local equilibrium quantities}
\label{leqs}

Under the assumption of local equilibrium, the macroscopic
quantities appearing in the hydrodynamic equations are given by
\begin{subequations}
\label{n0s}
\begin{equation}
\label{n0}
n = n_{+}\! -\! n_{-} = \frac{T^2}{v_g^2}
\frac{g_2\left(\mu_+/T\right)\!-\!g_2\left(-\mu_-/T\right)}
     {\left(1\!-\!u^2/v_g^2\right)^{3/2}},
\end{equation}
\begin{equation}
\label{nI0_0}
n_{I} \!=\! n_{+,0} \!+\! n_{-} \!=\! \frac{T^2}{v_g^2}
\frac{g_2\left(\mu_+/T\right)\!+\!g_2\left(-\mu_-/T\right)}
     {\left(1\!-\!u^2/v_g^2\right)^{3/2}},
\end{equation}
\end{subequations}
\begin{equation}
  \label{g2}
  g_2\left(\frac{\mu}{T}\right) = - \frac{N}{2\pi} {\rm Li}_2\left(-e^{\mu/T}\right),
\end{equation}
where ${\rm Li}_n$ is the polylogarithm,
\begin{equation}
\label{ne0}
n_{E} = 2\frac{T^3}{v_g^2}
\frac{1\!+\!u^2/(2v_g^2)}
     {\left(1\!-\!u^2/v_g^2\right)^{5/2}}
\left[g_3\left(\frac{\mu_+}{T}\right)\!+\!g_3\left(-\frac{\mu_-}{T}\right)\right]\!,
\end{equation}
\begin{equation}
\label{g3}
g_3\left(\frac{\mu}{T}\right) = - \frac{N}{2\pi} {\rm Li}_3\left(-e^{\mu/T}\right),
\end{equation}
\begin{subequations}
\label{hqsg}
\begin{equation}
\label{phyd}
P = n_{E} \frac{1-u^2/v_g^2}{2+u^2/v_g^2},
\end{equation}
\begin{equation}
\label{whyd}
W=n_{E}+P = \frac{3 n_{E}}{2+u^2/v_g^2},
\end{equation}
\begin{equation}
\bs{j} = n \bs{u},
\qquad
\bs{j}_{I} = n_{I} \bs{u},
\end{equation}
\begin{equation}
\label{jehyd}
\bs{j}_{E} = v_g^2 \bs{n}_{\bs{k}} = W\bs{u},
\end{equation}
\begin{equation}
\label{pihyd}
\Pi_{E}^{\alpha\beta} \!=\!
P\delta_{\alpha\beta} + v_g^{-2}Wu_\alpha u_\beta.
\end{equation}
\end{subequations}

\section{General expressions for the viscosity coefficients}
\label{visc}

The general expressions for the shear and Hall viscosities in
graphene
\begin{widetext}
\begin{subequations}
\label{etares}
\begin{equation}
\label{etaB0}
\eta
=
\frac{{\cal T}T}{4\alpha_g^2v_g^2}
\begin{pmatrix}
0 & 0 & 1
\end{pmatrix}
\widehat{\textswab{M}}_h
\left(1+\pi^2\gamma_B^2\widehat{\textswab{T}}_\eta^{-1}\widehat{\textswab{M}}_K
\widehat{\textswab{T}}_\eta^{-1}\widehat{\textswab{M}}_K
\right)^{\!\!-1}
\widehat{\textswab{T}}_\eta^{-1}
\begin{pmatrix}
\tilde{n} \cr
(x^2\!+\!\pi^2/3)/2 \cr
3\tilde{n}_{E}
\end{pmatrix}\!,
\end{equation}
\begin{equation}
\label{etaH0}
\eta_H 
=\pi\gamma_B
\frac{{\cal T}T}{4\alpha_g^2v_g^2}
\begin{pmatrix}
0 & 0 & 1
\end{pmatrix}
\widehat{\textswab{M}}_h
\left(1+\pi^2\gamma_B^2\widehat{\textswab{T}}_\eta^{-1}\widehat{\textswab{M}}_K
\widehat{\textswab{T}}_\eta^{-1}\widehat{\textswab{M}}_K
\right)^{\!\!-1}
\widehat{\textswab{T}}_\eta^{-1}\widehat{\textswab{M}}_K\widehat{\textswab{T}}_\eta^{-1}
\begin{pmatrix}
\tilde{n} \cr
(x^2\!+\!\pi^2/3)/2 \cr
3\tilde{n}_{E}
\end{pmatrix}\!,
\end{equation}
\end{subequations}
with the following notations
\begin{subequations}
\begin{equation}
\gamma_B = \frac{|e|v_g^2B}{\alpha_g^2cT^2},
\end{equation}
\begin{equation}
\label{t}
{\cal T} = T
\left[\ln\left(1+e^{\mu_+/T}\right)+\ln\left(1+e^{-\mu_-/T}\right)\right]
\qquad\rightarrow\qquad 
{\cal T} = 2T\ln2\cosh\frac{\mu}{2T},
\end{equation}
where the last expression is obtained in the limit $\mu_\pm=\mu$, that
is used hereafter,
\begin{equation}
\label{ntil}
\tilde{n}(x) = -{\rm Li}_2(-e^{x})\!+\!{\rm Li}_2(-e^{-x}),
\qquad
\tilde{n}_E(x) = -{\rm Li}_3(-e^{x})\!-\!{\rm Li}_3(-e^{-x}),
\qquad
x = \frac{\mu}{T},
\end{equation}
\begin{equation}
\widehat{\textswab{M}}_h=
\begin{pmatrix}
1 & \frac{xT}{\cal T} & 2\tilde{n} \frac{T}{\cal T} \cr
\frac{xT}{\cal T} & 1 & \left[x^2\!+\!\frac{\pi^2}{3}\right]\frac{T}{\cal T} \cr
2\tilde{n} \frac{T}{\cal T} & \left[x^2\!+\!\frac{\pi^2}{3}\right]\frac{T}{\cal T} &
6\tilde{n}_E \frac{T}{\cal T}
\end{pmatrix},
\end{equation}
\begin{equation}
\label{mk}
\widehat{\textswab{M}}_K
=
\begin{pmatrix}
\tanh\frac{x}{2} & 1 & \frac{\cal T}{T} \cr
1 & \tanh\frac{x}{2} & x \cr
\frac{\cal T}{T} & x & 2\tilde{n}
\end{pmatrix}
\end{equation}
\begin{equation}
\label{taueta}
\widehat{\textswab{T}}_\eta
=
\frac{2\pi}{\alpha_g^2}\frac{\cal T}{NT^2}
\begin{pmatrix}
\tilde\tau_{11}^{-1} & \tilde\tau_{12}^{-1} & \tilde\tau_{13}^{-1}\cr
\tilde\tau_{12}^{-1} & \tilde\tau_{22}^{-1} & \tilde\tau_{23}^{-1} \cr
\tilde\tau_{13}^{-1} & \tilde\tau_{23}^{-1} & \tilde\tau_{33}^{-1}
\end{pmatrix}.
\end{equation}
\end{subequations}
The ``scattering rates'' $\tilde\tau_{ij}^{-1}$ are obtained from the
collision integral within the three-mode approximation \cite{hydro1}
and are given by
\begin{equation}
\label{ttauij}
\frac{1}{\tilde\tau_{ij}} =  
(2\pi)^2 \alpha_g^2NT \left[\frac{NT}{v_g^2\partial n_0/\partial\mu}\right]
\int\frac{d^2Q}{(2\pi)^2}\frac{dW}{2\pi}
\frac{|\widetilde{U}|^2}{\sinh^2W}
\left[{Y}_{00}\widetilde{Y}_{ij}- \widetilde{Y}_{0j} \widetilde{Y}_{0i}\right]\!,
\end{equation}
where
\begin{subequations}
\label{tys}
  \begin{equation}
  \label{y00i}
  Y_{00}(\omega,\bs{q}) = \frac{1}{4\pi}
  \left[
    \frac{\theta(|\Omega|\leqslant1)}{\sqrt{1\!-\!\Omega^2}}\,
    {\cal Z}_0^>[I_1]
    +
    \frac{\theta(|\Omega|\geqslant1)}{\sqrt{\Omega^2\!-\!1}}\,
    {\cal Z}_0^<[I_1]
    \right]\!,
\end{equation}
\begin{equation}
  \label{ty_01}
\widetilde{Y}_{01}(\omega,\bs{q})
= -\frac{1}{\pi}\!
  \left[
    \theta(|\Omega|\leqslant1)\Omega\sqrt{1\!-\!\Omega^2}\,
    {\cal Z}^>_5[I]
    +
    \theta(|\Omega|\geqslant1)\Omega\sqrt{\Omega^2\!-\!1}\,
    {\cal Z}^<_5[I]
    \right],
\end{equation}
\begin{equation}
\label{ty_02}
\widetilde{Y}_{02}(\omega,\bs{q})
= \frac{1}{\pi}
\left[
\theta(|\Omega|\leqslant1)\Omega\sqrt{1\!-\!\Omega^2}\,
{\cal Z}^>_5[I_1]
-\frac{1}{2}
\theta(|\Omega|\geqslant1)\, {\rm sign}(\Omega) \sqrt{\Omega^2\!-\!1}\,
\widetilde{\cal Z}^<_4[I_1]
\right]
+
\frac{1}{4\pi}
\theta(|\Omega|\geqslant1)\frac{{\rm sign}(\Omega)}{\sqrt{\Omega^2\!-\!1}}\,{\cal Z}^<_0[I_1],
\end{equation}
\begin{eqnarray}
\label{ty_03}
&&
\widetilde{Y}_{03}(\omega,\bs{q})
=
Q\Omega \,Y_{00}(\omega,\bs{q})
+
\frac{1}{2\pi}\,Q\Omega
\left[
\theta(|\Omega|\leqslant1)\sqrt{1\!-\!\Omega^2}
\,{\cal Z}^>_3[I_1]
-
\theta(|\Omega|\geqslant1)\sqrt{\Omega^2\!-\!1}
\,{\cal Z}^<_3[I_1]
\right],
\end{eqnarray}
\begin{equation}
\label{ty_11}
\widetilde{Y}_{11}(\omega,\bs{q})
= \frac{1}{\pi}\!
\left[
\theta(|\Omega|\leqslant1)\sqrt{1\!-\!\Omega^2}\,{\cal Z}^>_4[I_1]
+
\theta(|\Omega|\geqslant1)\sqrt{\Omega^2\!-\!1}\,{\cal Z}^<_4[I_1]
\right]\!,
\end{equation}
\begin{equation}
\label{ty_12}
\widetilde{Y}_{12}(\omega,\bs{q})
= -\frac{1}{\pi}
\theta(|\Omega|\leqslant1)\sqrt{1\!-\!\Omega^2}
\,{\cal Z}^>_4[I],
\end{equation}
\begin{equation}
\label{ty_13}
\widetilde{Y}_{13}(\omega,\bs{q})
= -\frac{Q}{\pi}
\left[
\theta(|\Omega|\leqslant1)\sqrt{1\!-\!\Omega^2}\,{\cal Z}^>_5[I]
+
\theta(|\Omega|\geqslant1)\sqrt{\Omega^2\!-\!1}\,{\cal Z}^<_5[I]
\right]\!,
\end{equation}
\begin{eqnarray}
\label{ty_22}
\widetilde{Y}_{22}(\omega,\bs{q})
= \frac{1}{\pi}\left[
\theta(|\Omega|\leqslant1)\sqrt{1\!-\!\Omega^2}
\,{\cal Z}^>_4[I_1]
-
\theta(|\Omega|\geqslant1)\sqrt{\Omega^2\!-\!1}
\,{\cal Z}^<_4[I_1]
+ \frac{1}{4}
\frac{\theta(|\Omega|\geqslant1)}{\sqrt{\Omega^2\!-\!1}}
\,{\cal Z}^<_0[I_1]
\right],
\end{eqnarray}
\begin{eqnarray}
\label{ty_23}
\widetilde{Y}_{23}(\omega,\bs{q})
= \frac{Q}{\pi}\left[
\theta(|\Omega|\leqslant1)\sqrt{1\!-\!\Omega^2}
\,{\cal Z}^>_5[I_1]
-
\theta(|\Omega|\geqslant1)|\Omega|\sqrt{\Omega^2\!-\!1}
\,{\cal Z}^<_4[I_1]
+ \frac{|\Omega|}{4}
\frac{\theta(|\Omega|\geqslant1)}{\sqrt{\Omega^2\!-\!1}}
\,{\cal Z}^<_0[I_1]
\right]\!,
\end{eqnarray}
\begin{eqnarray}
\label{ty_33}
\widetilde{Y}_{33}(\omega,\bs{q})
=
\frac{Q^2}{\pi}\left[
\theta(|\Omega|\leqslant1)\sqrt{1\!-\!\Omega^2}
\,{\cal Z}^>_3[I_1]
-
\theta(|\Omega|\geqslant1)\sqrt{\Omega^2\!-\!1}
\,{\cal Z}^<_3[I_1]\right]
+Q^2\Omega^2 Y_{00}(\omega,\bs{q}).
\end{eqnarray}
\end{subequations}
These functions are expressed in terms of the integrals
\begin{subequations}
\label{Zints}
\begin{equation}
\label{z0}
{\cal Z}^>_0[I] = \int\limits_1^\infty dz\sqrt{z^2\!-\!1} \, I(z),
\qquad
{\cal Z}^<_0[I] = \int\limits_0^1 dz\sqrt{1\!-\!z^2} \, I(z),
\end{equation}
\begin{equation}
\label{z3}
{\cal Z}^>_3[I] = \int\limits_1^\infty dz \frac{\left(z^2\!-\!1\right)^{3/2}}{z^2\!-\!\Omega^2} \, I(z),
\qquad
{\cal Z}^<_3[I] = \int\limits_0^1 dz \frac{\left(1\!-\!z^2\right)^{3/2}}{\Omega^2\!-\!z^2}\, I(z).
\end{equation}
\begin{equation}
\label{z4}
{\cal Z}^>_4[I] = \int\limits_1^\infty dz
\frac{\left(z^2\!-\!1\right)^{3/2}}{\left(z^2\!-\!\Omega^2\right)^2} \, I(z),
\qquad
{\cal Z}^<_4[I] = \int\limits_0^1 dz
\frac{\left(1\!-\!z^2\right)^{3/2}}{\left(\Omega^2\!-\!z^2\right)^2}\, I(z),
\qquad
\widetilde{\cal Z}^<_4[I] = \int\limits_0^1 dz
\frac{\left(1\!-\!z^2\right)^{3/2}}{\left(\Omega^2\!-\!z^2\right)^2}
\left(\Omega^2\!+\!z^2\right)\, I(z),
\end{equation}
\begin{equation}
\label{z5}
{\cal Z}^>_5[I] = \int\limits_1^\infty dz
\frac{z\left(z^2\!-\!1\right)^{3/2}}{\left(z^2\!-\!\Omega^2\right)^2} \, I(z),
\qquad
{\cal Z}^<_5[I] = \int\limits_0^1 dz
\frac{z\left(1\!-\!z^2\right)^{3/2}}{\left(\Omega^2\!-\!z^2\right)^2}\, I(z),
\end{equation}
\end{subequations}
that are evaluated for either of the two functions
\begin{subequations}
\label{izs}
\begin{equation}
\label{i_10}
I_1(z) = \tanh\frac{zQ+W+x}{2}+\tanh\frac{zQ+W-x}{2}-\tanh\frac{zQ-W+x}{2}-\tanh\frac{zQ-W-x}{2},
\end{equation}
\begin{equation}
\label{iz0}
I(z) = \tanh\frac{zQ+W+x}{2}-\tanh\frac{zQ+W-x}{2}-\tanh\frac{zQ-W+x}{2}+\tanh\frac{zQ-W-x}{2}.
\end{equation}
\end{subequations}
\end{widetext}
The frequency and momentum are expressed in terms of the
dimensionless variables
\begin{equation}
\label{dv}
\bs{Q} = \frac{v_g\bs{q}}{2T},
\qquad
W = \frac{\omega}{2T}.
\end{equation}
Finally, the Coulomb interaction has the form
\begin{equation}
\label{coulomb}
U(\omega,\bs{q})= \frac{2\pi e^2}{q} \widetilde{U} = \frac{2\pi\alpha_g v_g}{q}\widetilde{U}, 
\qquad
\alpha_g=\frac{e^2}{v_g\varepsilon},
\end{equation}
where $\varepsilon$ is the dielectric constant of the dielectric
environment and the dimensionless factor $\widetilde{U}$ accounts for
screening.

\bibliography{viscosity_refs}

\end{document}